\documentclass[11pt,twoside,a4paper]{article}

\usepackage{graphicx}

\newcommand{\Tr}{\makebox{ Tr }}

\newcommand{\GeV}{\makebox{ GeV}}

\newcommand{\fm}{\makebox{ fm}}

\newcommand{\beq}{\begin{equation}}

\newcommand{\enq}{\end{equation}}

\newcommand{\beqa}{\begin{eqnarray}}

\newcommand{\enqa}{\end{eqnarray}}

\newcommand{\nn}{\nonumber}

\newcommand{\labelm}[1]{\label{#1}}

\newcommand{\lbfi}[1]{\labelm{#1}\end{figure}}

\newcommand{\lbq}[1]{\labelm{#1}\enq}

\newcommand{\lbqa}[1]{\labelm{#1}\enqa}

\newcommand{\befi}[2]{\begin{figure}[ht] \leavevmode \begin{center} \includegraphics[width=#2]{#1.eps} \end{center}}

\newcommand{\betwofi}[5]{
\begin{figure}[ht]
\leavevmode
\begin{minipage}{#5}
\includegraphics[width=#5]{#1.eps}
\end{minipage}
\hfill
\begin{minipage}{#5}
\includegraphics[width=#5]{#3.eps}
\end{minipage}\\
\begin{minipage}{#5}
\begin{center}
\small #2
\end{center}
\end{minipage}
\hfill
\begin{minipage}{#5}
\begin{center}
\small #4
\end{center}
\end{minipage}
}

\newcommand{\eq}[1]{eq.(\ref{#1})}

\newcommand{\fig}[1]{fig.(\ref{#1})}

\newcommand{\lbcap}[3]{\begin{center}\begin{minipage}{#1}\caption{\small #2}\labelm{#3}\end{minipage}\end{center}\end{figure}}

\newcommand{\lbtab}[3]{\centering\begin{minipage}{#1}\caption{\small #2}\labelm{#3}\end{minipage}\end{table}}

\newcommand{\pa}{\partial}

\newcommand{\cF}{\mbox{$\cal F$}}

\newcommand{\cS}{\mbox{$\cal S$}}

\newcommand{\cN}{\mbox{$\cal N$}}

\newcommand{\bA}{\mbox{\bf A}}

\newcommand{\bW}{\mbox{\bf W}}

\newcommand{\bo}{\mbox{\bf 1}}

\newcommand{\al}{\alpha}

\newcommand{\ga}{\gamma}

\newcommand{\de}{\delta}

\newcommand{\ep}{\epsilon}

\renewcommand{\th}{\theta}

\newcommand{\ka}{\kappa}

\newcommand{\la}{\lambda}

\newcommand{\rh}{\rho}

\newcommand{\si}{\sigma}

\newcommand{\ph}{\phi}

\newcommand{\ch}{\chi}

\newcommand{\om}{\omega}

\newcommand{\De}{\Delta}

\newcommand{\La}{\Lambda}

\newcommand{\Ps}{\Psi}

\begin{document}
\title{\begin{flushright}\normalsize
TAUP-2507-98\end{flushright}
\vspace*{1cm}\LARGE \bf Energy and $Q^2$ dependence of elastic vectormeson production and the proton structure function $F_2$}
\author{
{\LARGE Michael Rueter}\thanks{supported by a MINERVA-fellowship}\\[1cm]
\itshape School of Physics and Astronomy\\
\itshape Department of High Energy Physics\\
\itshape Tel-Aviv University\\
\itshape 69978 Tel-Aviv, Israel\\
\itshape e-mail: {\tt rueter@post.tau.ac.il}}
\date{}
\maketitle
\thispagestyle{empty}
\begin{abstract}
  
  In the framework of the Model of the Stochastic Vacuum elastic
  hadron-hadron scattering, photo- and electroproduction of
  vectormesons and also $F_2(Q^2)$ can be well described at center of
  mass energy approximately 20 GeV. The scattering amplitude is
  derived by smearing the color dipole-dipole scattering, which is
  calculated nonperturbatively in the Model of the Stochastic Vacuum,
  with appropriate wavefunctions. For the considered processes the
  dipoles have extensions in the range of hadron sizes. We now extend
  this idea to small dipoles and high energies. The energy dependence
  is modeled in a phenomenological way: we assume that there a two
  pomerons, the soft- and the hard-pomeron, each being a simple pole
  in the complex angular plane. We couple dipoles of hadronic size to
  the soft-pomeron and small dipoles to the hard-pomeron. For small
  dipoles we take the perturbative gluon exchange into account. By
  that way we obtain an energy dependent dipole-dipole scattering
  amplitude which can be used for all the processes with the same
  parameters. We show that this approach can describe in addition to
  all the low energy results (20 GeV) also the HERA data for the
  considered processes in a large energy and $Q^2$ range. Especially
  the right transition from the soft to the hard behavior is observed.

\end{abstract}
\newpage
\setcounter{page}{1}
\section{Introduction}
One of the most exciting results of HERA is the observed very
different energy behavior of \mbox{$\ga^*$-$p$} scattering depending
on the virtuality of the photon and the considered final state. There
are now many data for elastic photo- and electroproduction of
vectormesons and the proton structure function $F_2(x,Q^2)$ in a large
kinematic regime which allow to study this subject in detail.

In most of the theoretical approaches these processes are studied by
first calculating the fluctuation of the photon into a color neutral
quark-antiquark pair, a so called color dipole. This color dipole then
interacts with the proton. Depending on the virtuality of the photon
and on the final state of the reaction one can vary the size of the
dipoles mainly involved in the interaction. For example by increasing
$Q^2$ the dipoles contributing to $F_2$ or to electroproduction of
vectormesons become smaller.

The energy dependence of the total cross sections of hadronic
interactions can be well described by the Donnachie-Landshoff
parameterization \cite{Donnachie:1992}, that is by the soft-pomeron
exchange based on Regge theory \cite{Collins:Buch}. But the HERA
results show that processes with small dipoles involved rise much
stronger with the energy, as can be seen for example from $J/\Ps$
production or from the behavior of $F_2$ for small $x$ and large
$Q^2$. The interaction responsible for this strong rise was called
hard-pomeron exchange. Usually the calculations for large $Q^2$ and
small $x$ are based on perturbative QCD like the BFKL pomeron
\cite{Balitskii:1978,Kuraev:1976} or the DGLAP evolution
\cite{Altarelli:1977,Gribov:1972,Lipatov:1975,Dokshitser:1977}. For
the BFKL approach to $F_2$ see for example references
\cite{Nikolaev:1994III,Nikolaev:1994IIII,Nikolaev:1997} and for the
DGLAP approach references \cite{Gluck:1989,Gluck:1992,Gluck:1995}. But
if $x$ is small enough, $W$, the internal cm-energy is still the
biggest scale and thus Regge theory should be applicable.

There are some approaches to describe the transition from the soft-
to the hard-pomeron behavior. One idea is that the hard-pomeron is
always present and the transition to the soft behavior is due to
shadowing effects \cite{Gotsman:1994,Gotsman:1996}. Another
possibility is to vary the pomeron intercept with $Q^2$
\cite{Capella:1994,Bertini:1995}. The behavior of the proton structure
function at small $x$ and small or moderate $Q^2$ can be described
\cite{Gotsman:1997,Desgrolard:1998II,Martin:1998} by splitting the
dipole-proton scattering in a perturbative and nonperturbative regime
where the involved dipoles are small or large respectively. For the
small dipoles the perturbative QCD methods (DGLAP) are used. For the
nonperturbative regime one uses vectormeson dominance which allows, by
using the additive quark model, to connect the scattering of large
dipoles with the proton with the Regge behavior of the measured total
hadronic cross sections. A similar approach based on BFKL exists for
the electroproduction of vectormesons
\cite{Nemchik:1996II,Nemchik:1997}.

In our approach, which we use to describe $F_2$ and the vectormeson
production simultaneously, we assume that there are two pomerons, the
soft- and the hard-pomeron, each being a simple pole in the complex
angular plane. We then make a phenomenological ansatz for the coupling
of these pomerons to the color dipoles in such a way that the
hard-pomeron couples to small and the soft-pomeron to large dipoles.
The different energy behavior of the considered processes is due to a
drastically change of the relative weight of the two pomeron
contributions because of the different sizes of the dipoles involved
in the scattering processes. This scheme is very similar to a recent
publication of Donnachie and Landshoff \cite{Donnachie:1998}. In this
paper DL showed that by fitting the hard-pomeron intercept the $F_2$
data for not too large $x$ can be described by the two pomerons and
the leading Regge-trajectory. In contrary to our work they also had to
fit the coupling of the pomerons as a function of $Q^2$.

The building block of our calculation is the dipole-dipole scattering
amplitude. It is calculated in the framework of the Model of the
Stochastic Vacuum (MSV) \cite{Dosch:1987,Dosch:1988}. Within this
framework elastic hadron-hadron scattering
\cite{Dosch:1994,Rueter:1996,Rueter:1996III}, hadron-dipole scattering
\cite{Rueter:1997II}, photo- and electroproduction of vectormesons
\cite{Dosch:1997,Kulzinger:1998} and $\pi^0$ \cite{Rueter:1998} and
the proton structure function $F_2$ \cite{Dosch:1997II} were
calculated but the cm-energy was always fixed at $20 \GeV$. We now
extend this approach to higher energies as described above. If we
consider processes where one of the dipoles is very small we include
in our approach also the leading perturbative gluon exchange.

Our paper is organized as follows: In section 2 we review our
calculation of the dipole-proton scattering amplitude within the MSV.
In section 3 we describe in detail how the two pomerons and their
coupling to the dipoles is in-cooperated in our model. We also
calculate the leading perturbative contribution for very small
dipoles. In section 4 we present our results for the different
reactions and close with a summary in section 5. For some technical
steps we append an appendix.
\section{Review of our approach}
All our previous applications of the MSV on high-energy scattering are
based on dipole-dipole scattering smeared with appropriate
wavefunctions. In this letter we do not derive the dipole-dipole
result but refer to the literature and reviews
\cite{Dosch:1994,Dosch:1997III,Nachtmann:1996}. In the remaining
section we follow the very recent paper \cite{Rueter:1998}.

The soft high-energy scattering is calculated first using an eikonal approximation in a fixed gluon background field \cite{Nachtmann:1991}. The local gauge invariant color dipoles are represented in space-time as Wegner-Wilson loops $\bW[\cS]=P\exp\left[-ig \oint_{\partial S} \bA_\mu
(z)\ dz^\mu\right]$
whose lightlike sides are formed by the quark and antiquark
pathes, and front ends by the Schwinger strings ensuring local gauge
invariance (see \fig{vac2}).
\befi{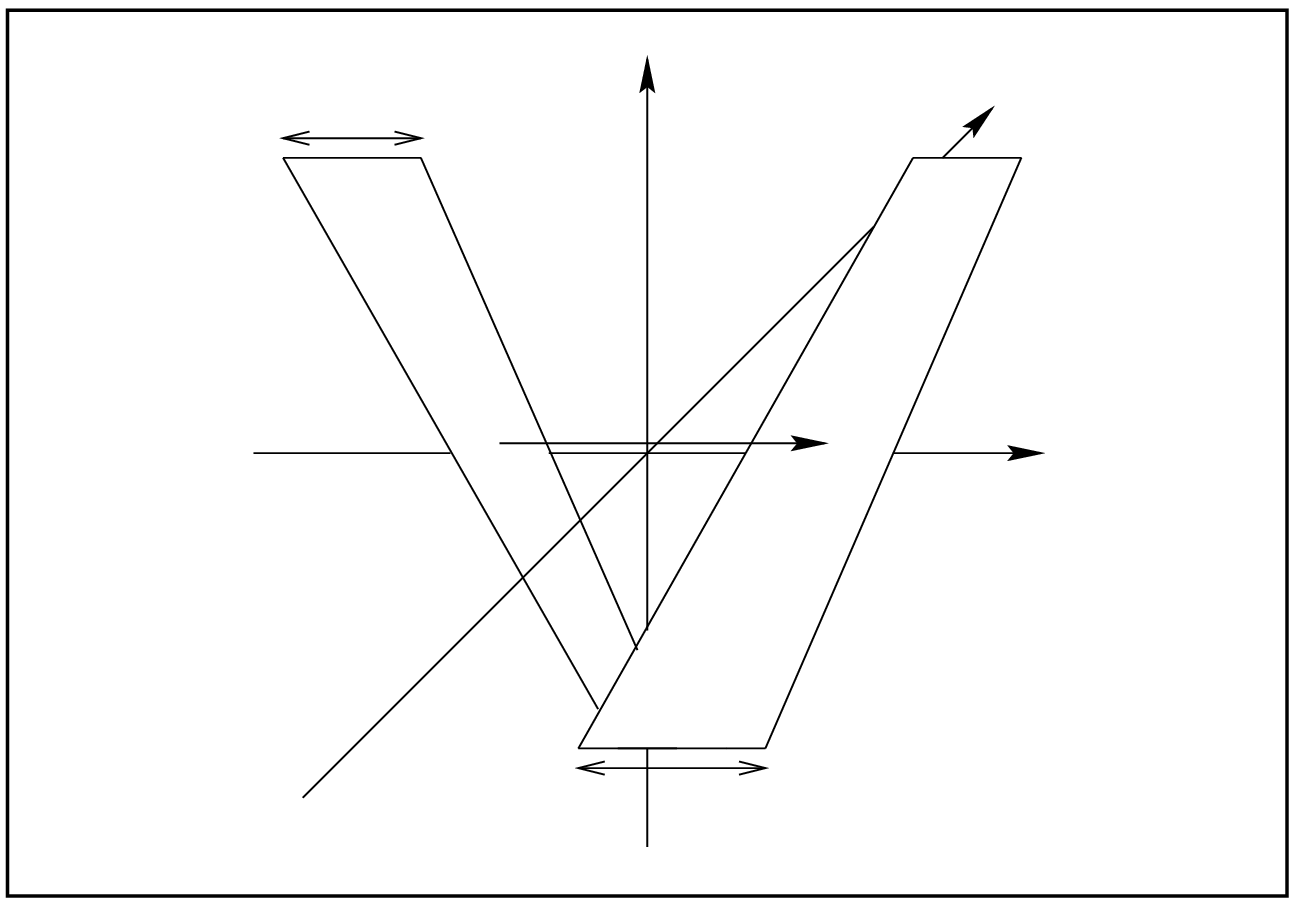}{8.5cm}
\unitlength.85cm
\begin{picture}(0,0)
\small
\put(5.4,5.4){loop 2}
\put(12.4,5.4){loop 1}
\put(12.2,3.9){$\vec{x}$}
\put(9.5,7.1){$x^0$}
\put(12.2,6.9){$x^3$}
\put(6.9,7.0){$\vec{R}_2$}
\put(9.5,1.4){$\vec{R}_1$}
\put(9.0,4.6){$\vec{b}$}
\end{picture}
\lbcap{14cm}{Wegner-Wilson loops formed by the paths of quarks
and antiquarks inside two dipoles. The impact parameter $\vec b$ is the
distance vector between the middle lines of the two loops. $\vec R_1$ and
$\vec R_2$ are the vectors in the transverse plane from the quark
lines to the antiquark lines of dipole 1 and 2 respectively. The front lines
of the loops guarantee that the dipoles behave as singlets under local
gauge transformations.}{vac2}

The resulting loop-loop amplitude is not only specified by the
impact parameter, but also by the transverse extension vectors of the
loops. The dipole-dipole profile function is then obtained by integrating over the gluon background field:
\beq
\tilde J (\vec b,\vec R_1,\vec R_2)=
\frac{-\big< W_1 W_2\big> _A}{ \big< {1\over N_{\rm C}} \Tr \bW_1(0,\vec R_1)\big> _A
\big< {1\over N_{\rm C}}\Tr\bW_2(0,\vec R_2)\big>_A},
\lbq{profilewithloops}
where the brackets denote functional integration over the background field $\bA$. The path $\partial S_1$ of the closed Wegner-Wilson loop $\bW[\cS_{1}]$ in
\beq
W_i=\frac{1}{N_{\rm C}}\Tr \left\{\bW[\cS_{i}]-\bo\right\}
\lbq{WWloopb}
is a rectangle whose long sides are formed by the quark path
$\Gamma_1^q=(x^0,\vec b/2+\vec R_1/2,x^3=x^0)$ and the antiquark path
$\Gamma_1^q=(x^0,\vec b/2-\vec R_1/2,x^3=x^0)$ and whose front sides are
formed by lines from $(T,\vec b/2+\vec R_1/2,T)$ to $(T,\vec b/2-\vec
R_1/2,T)$ for large positive and negative $T$ (we will then take the
limit $T\to \infty$). $W_2$ is constructed
analogously. The denominator in \eq{profilewithloops} is the loop
renormalization.

By expanding the exponentials of the Wegner-Wilson loops and using the Gaussian approximation adopted in the MSV we can express the dipole-dipole profile function (\eq{profilewithloops}) as a product of nonlocal gluon condensates. For this condensate we make a nonperturbative ansatz in agreement with lattice measurements of this quantity \cite{DiGiacomo:1992,DiGiacomo:1996}. It was shown that the leading contribution to the dipole-dipole profile function is even under charge parity, like the pomeron and two gluon exchange, and is given by
\beq
\tilde{J}=\frac{1}{8 N_{\rm C}^2(N_{\rm C}^2-1)12^2}\tilde\ch^2
\lbq{ddcplus}
where
\[
\tilde\ch^2=\left( \tilde{\ch}_{11}-\tilde{\ch}_{12}-\tilde{\ch}_{21}+\tilde{\ch}_{22}\right)^2.
\]
The real functions $\tilde{\ch}_{ij}$ depend only on the transversal coordinates and are given by \cite{Dosch:1994,Rueter:1997II}:
\beqa
\tilde{\ch}_{ij}&=&\langle g^2FF\rangle\left( \ka \, \int_0^1dw_1\, \int_0^1dw_2\, \vec{r}_{1i}\cdot\vec{r}_{2j}\, f_1\left( |w_1 \vec{r}_{1i}-w_2\vec{r}_{2j}|\right)\right.\nn\\
&&\hspace{2cm}+ (1-\ka )f_2\left( |\vec{r}_{1i}-\vec{r}_{2j}|\right)\bigg).
\lbqa{chi}
The vector $\vec{r}_{1i}$ ($\vec{r}_{2j}$) points to constituent $i$ ($j$) of dipole 1 (2) and is a function of $\vec{b}$, $\vec{R}$ and $z$ as indicated in \fig{trans}. The usual gluon condensate is denoted by $\langle g^2FF\rangle$ and the parameter $\ka$ and the two functions $f_1$ and $f_2$ depend on the explicit ansatz for the nonlocal gluon condensate and fall off on the length scale given by the correlation length $a$. Their explicit form is given in reference \cite{Rueter:1997II}. In reference \cite{Dosch:1994} it was also shown that one of the $w$-integrations in \eq{chi} can be done analytically.
\befi{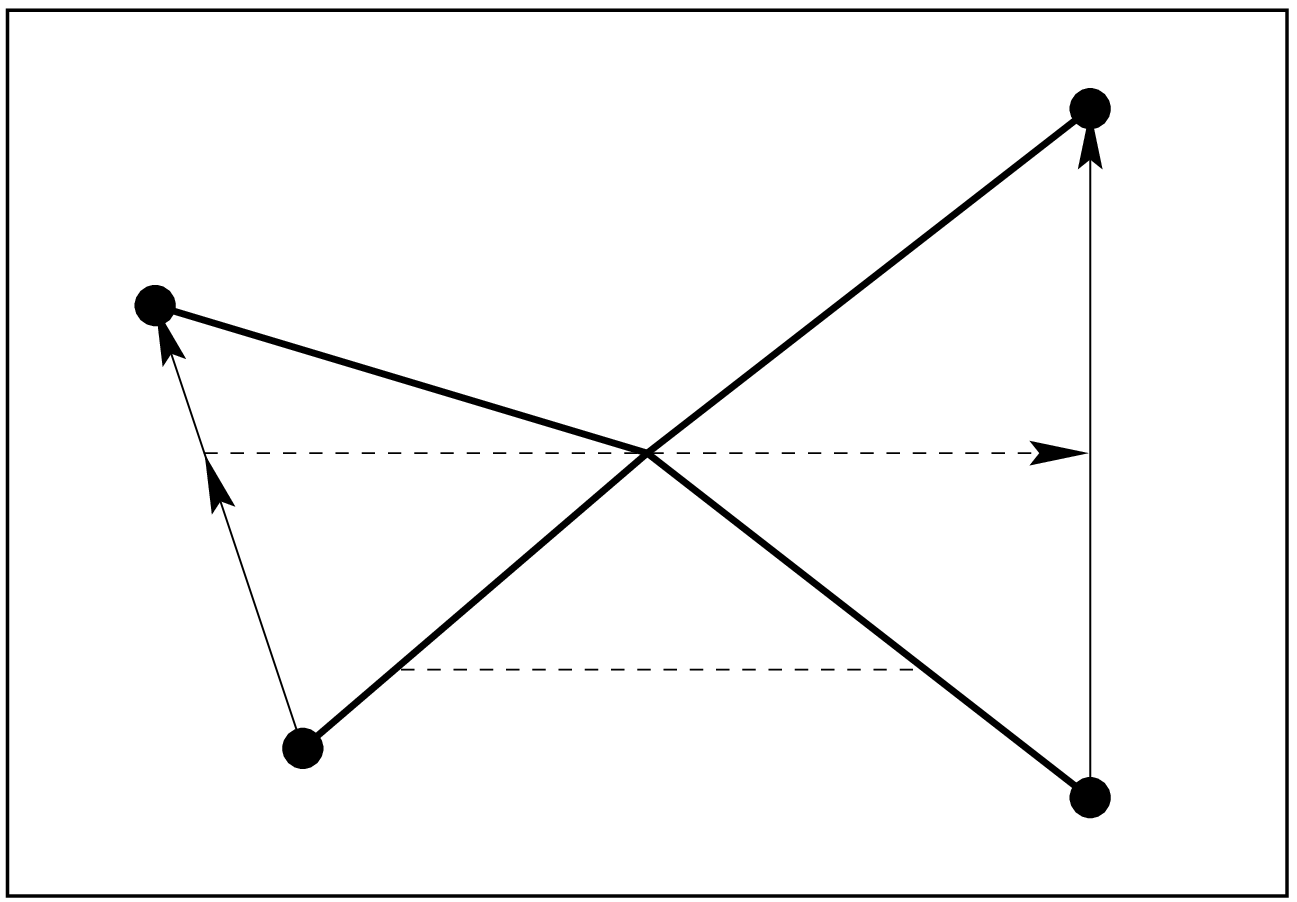}{8cm}
\unitlength1cm
\begin{picture}(0,0)
\small
\put(5.7,1.3){1}
\put(4.8,4.7){2}
\put(10.6,1.0){1}
\put(10.6,6.0){2}
\put(6.5,3.1){$\vec b$}
\put(4.2,3.7){$z\vec R_1$}
\put(4.7,2.5){$\bar z \vec R_1$}
\put(10.9,2.5){$\vec R_2$}
\put(7.7,1.8){$\tilde\ch _{11}$}
\end{picture}
\lbcap{14cm}{A geometrical picture of the scattering in the transversal plane. The constituents are denoted by the black dots. The two dipoles scatter of with impact parameter $\vec{b}$. The thick lines from the central point to the (anti-)quarks denote the paths covered by the integration in \eq{chi}. The term $\tilde\chi_{ij}$ represents the contribution of
a correlator of a field strength on the piece $i$ of dipole 1 with a
field strength $j$ of dipole 2. The integration has to be performed
over all the transversal projections of the surface, i.e.~1 and 2 of
dipole 1 combined with 1 and 2 of dipole 2. The impact parameter $\vec
b$ points to the lightcone barycenter of the dipoles, i.e.~the
distance between the quark and antiquark is divided according to the
longitudinal momentum fraction of each constituent which is given by $z$ and $\bar{z}=1-z$ \protect\cite{Dosch:1997}.}{trans}

By smearing the dipole-dipole profile function with appropriate wavefunctions we obtain the process dependent profile function $J$. To obtain the scattering amplitude at center of mass energy $s$ and momentum transfer \mbox{$t=-\vec{\De}_\perp^2$} one has to integrate over the impact parameter $\vec{b}$
\beq
T(s,t) = 2is \int \, d^2b\,e^{-i\vec{\De}_\perp\cdot\vec{b}}\,J.
\lbq{T}
For the total cross section follows
\beq
\si^{\rm tot}=\frac{1}{s}{\rm Im}T(s,0)
\lbq{sigtotundslope}
which is independent of the center of mass energy $s$. For the differential cross section we obtain
\beq
\frac{d\si^{\rm el}}{dt}=\frac{1}{16\pi s^2}|T|^2.
\lbq{dsigdt}
As mentioned these results are independent of the energy $s$. The parameters of the MSV were fixed for $p$-$\bar p$ scattering at $\sqrt s =20 \GeV$ \cite{Dosch:1994}. Using the most general ansatz of the MSV the parameters change slightly as calculated in \cite{Rueter:1997} and published in \cite{Dosch:1997}: 
\beq
a=0.346\fm,\;\langle g^2FF\rangle = 2.49\GeV^4,\;\ka= 0.74,\;S_{\rm P}=0.739\fm,
\lbq{MSVpara}
where $S_{\rm P}$ is the proton size. Within this framework elastic hadron-hadron scattering \cite{Dosch:1994,Rueter:1996,Rueter:1996III}, hadron-dipole scattering \cite{Rueter:1997II}, photo- and electroproduction of vectormesons \cite{Dosch:1997,Kulzinger:1998} and $\pi^0$ \cite{Rueter:1998} and the proton structure function $F_2$ \cite{Dosch:1997II} were calculated. In all these references we were limited to a cm-energy of about 20 GeV and dipoles of hadronic size (thus for photons the virtuality was limited). For very small dipoles this approach has to be modified because then a perturbative calculation has to replace the nonperturbative Model of the Stochastic Vacuum.

In the next section we introduce an energy dependence in a
phenomenological way: we assume that there are two pomerons each being
a simple pole in the complex angular plane. The coupling of these
pomerons will be modeled in such a way that the scattering of two
dipoles of hadronic size is due to soft-pomeron exchange whereas small
dipoles couple to the hard-pomeron. We also extend our approach to
very small dipoles by taking the perturbative contribution to
\eq{profilewithloops} into account.
\section{Extension to very small dipoles and high energies}
In this section we extend our model to very small dipoles and high
energies. Our approach is based on the dipole-dipole scattering. The
experimental data seem to indicate that the energy dependence must be
very different for the scattering of small and large dipoles. In this
paper we model this transition from small to large dipoles in a
phenomenological way and introduce an effective dipole-dipole
interaction: The energy dependence is put {\it by hand}. We assume
that there are two pomerons which are simple poles in the complex
angular plane. The coupling to the dipoles is modeled in such a way
that for small dipoles the hard-pomeron and for large dipoles the
soft-pomeron gives the main contribution to the scattering process.
The cut between the soft- and hard-pomeron will be given by $c$. In
addition we have to switch off the contributions calculated in the
nonperturbative MSV if one of the dipoles is very small. Therefor we
introduce a second cut $r_{\rm cut}$. The nonperturbative interaction
of large dipoles, larger than $r_{\rm cut}$, is calculated using the
Model of the Stochastic Vacuum. With this modification we can already
describe very well the experimental data of vectormeson production and
$F_2$ for not too large $Q^2\le 35\GeV^2$. If we want to extend the
approach to even harder processes, which is not the main goal off this
paper, we have to calculate the interaction of dipoles smaller than
$r_{\rm cut}$ perturbatively. For simplicity we will use only the
leading perturbative contribution, the two gluon exchange. The two
cuts, $c$ and $r_{\rm cut}$, are the two important new parameters in
our model. To implement these scheme in our model we proceed as
follows:

We begin with the two pomerons. As already mentioned our idea is the
following: for physical processes that involve large dipoles,
especially elastic hadron-hadron scattering, we want to obtain the
soft-pomeron behavior, that is $\si_{\rm tot}\propto \sqrt
s^{\,0.16}$. For processes which are dominated by small dipoles we
want to obtain the hard-pomeron. Here we take the proton structure
function $F_2(x,Q^2)$ at $Q^2\approx 20 \GeV^2$ as a function of $x$
for small $x$ as measured at HERA \cite{Aid:1996III,Derrick:1996IIII}.
Off course there exist also data for much larger $Q^2$ but to separate
the hard- from the soft-pomeron $W$ has to be very large that is $x$
very small. For $Q^2>>20\GeV^2$ this kinematic regime ($x\le 0.01$) is
still not covered so well experimentally. For $Q^2=20\GeV^2$ the data
are well described by $F_2\propto W^{0.56}$
\cite{Aid:1996III,Adloff:1997}, where $W^2=Q^2/x-Q^2+m_{\rm P}^2$. We
will not fit this hard-pomeron power but take it as it is. To consider
photo- and electroproduction of vectormesons we need also the $t$
dependence of the pomerons. For the slope of the soft-pomeron we take
$0.25\GeV^{-2}$, the value obtained by Donnachie and Landshoff. There
is evidence that the slope of the hard-pomeron is quite small because
the experimental data indicate that there is no shrinkage of the
$B$-slope for photoproduction of $J/\Ps$ \cite{Levy:1997}. In this
paper we assume the slope of the hard-pomeron to be zero at least for
$|t|\le 0.5\GeV^2$. To take the energy dependence of the two pomerons
into account we replace the dipole-dipole profile function
(\eq{ddcplus}) integrated over the impact parameter with
\beqa
&&\int \, d^2b\,e^{-i\vec{\De}_\perp\cdot\vec{b}}\,\frac{1}{8 N_{\rm C}^2(N_{\rm C}^2-1)12^2}\tilde{\ch}^2\left( \vec R_1, \vec R_2\right)\label{Jtildetwopomeron}\\
&&\times\left(f_{\rm h}(R_1,R_2)\left(\frac{W}{20\GeV}\right)^{0.56}+f_{\rm s}(R_1,R_2)\left(\frac{W}{20\GeV}\right)^{2(0.08+0.25{\rm GeV}^{-2}t)}\right)\nn
\end{eqnarray}
where $f_{\rm h}(R_1,R_2)$ and $f_{\rm s}(R_1,R_2)$ are the couplings of the hard- or soft-pomeron to dipoles of given size. Here $W$ denotes the internal energy and is $\sqrt s$ for elastic hadron-hadron scattering. Due to the experiments we assume that $f_{\rm h}$ has to vanish for two large dipoles whereas $f_{\rm s}$ has to vanish if at least one dipole is very small. In this paper we make the most simple ansatz for these couplings by introducing only one parameter, the cut $c$ between the two pomerons:
\beqa
f_{\rm s}(R_1,R_2)&=&\left\{ {1\; | \; R_1 \;{\rm and}\; R_2> c} \atop{0\;| \;{\rm else}}\right\}\nn\\
f_{\rm h}(R_1,R_2)&=&\left\{ {1\; | \; R_1 {\rm \; or \;} R_2< c} \atop{0\;| \;{\rm else}}\right\}.
\lbqa{pomeroncoupling}
This {\it hard} cut between the soft- and hard-pomeron at the scale $c$ is off course an oversimplification of the {\it real physics}. If we calculate in this framework the cross section of dipole-dipole scattering we obtain for dipoles smaller then $c$ only the hard- and for dipoles larger than $c$ only the soft-pomeron contributions. In a more realistic model one would expect a smooth transition, that is $f_{\rm h/s}(R_1,R_2)$ being smooth functions. However, it will turn out that our very simple ansatz can describe the data very well which shows that only the scale of the cut is important and not the explicit form of the couplings. We want also mention that the energy behavior of the scattering of two small dipoles is yet not tested experimentally and $\ga^*$-$\ga^*$ scattering will be a very interesting probe for this issue. In this paper one dipole is always big because we only look at elastic proton scattering.

To investigate physical processes we then smear the energy dependent
dipole-dipole profile function (\eq{Jtildetwopomeron}) with
appropriate wavefunctions. By this way we obtain to all processes
contributions from the two pomerons but the relative weight of them
will depend strongly on the wavefunctions. Here we have to make an
important remark: The energy dependence of the scattering amplitude of
physical processes can only be written as
\[
a\,W^{2(1.08+0.25{\rm GeV}^{-2}t)}+b\, W^{2*1.28}
\]
if the wavefunctions are independent of the energy $W$. This will be
not the case for photo- and electroproduction of vectormesons or for
$F_2$. The reason is that we have to introduce an energy dependence of
the photon wavefunction to ensure energy conservation and the validity
of the eikonal approximation adopted in our model. To do so we have to
cut the end-points of the wavefunction for small $W$ (for more details
see the next section). The end-points of the wavefunctions are
especially important for large values of $Q^2$ and the cutting
diminishes the cross sections. For asymptotic large $W$, that is very
small $x$ this cutting has no effect. So this energy dependence of the
photon wavefunction makes the effective energy dependence of $F_2$ for
very large $Q^2>20\GeV^2$ at intermediate $x$ stronger than $W^{0.56}$
in agreement with the experiment.

For elastic hadron-hadron scattering the dipole sizes are larger than
the cut $c$ and thus only the soft-pomeron contributes. This is true
for all the processes that we have investigated in the past. Also the
cm-energy was limited to 20 GeV and thus the results of this processes
are unchanged. This allows us to take the old values for the
parameters of the MSV (\eq{MSVpara}). For all these processes the
energy dependence will by given by the soft-pomeron. Smearing the new
dipole-dipole profile function with {\it hard} wavefunctions, that is
with very small mean size, we obtain a profile function proportional
to $W^{0.56}$. But increasing the mean size of the dipoles we obtain a
soft transition of the effective behavior from the hard- to the
soft-pomeron because the relative weight of them changes drastically.
This transition will be seen for $F_2$ and for electroproduction of
vectormesons by varying $Q^2$, where we have for the $J/\Ps$ already
for $Q^2=0$ contributions from the hard-pomeron resulting in a strong
rise with $W$ as measured at HERA.

The contributions of the hard-pomeron are important if at least one
dipole is small. But for very small dipoles the dipole-dipole profile
function has to be calculated perturbatively. The leading perturbative
contribution to the dipole-dipole interaction (\eq{profilewithloops})
is easily calculated \cite{Rueter:1997} resulting in an additional
contribution to \eq{ddcplus} from the two gluon exchange. Off course
the two gluon exchange can only be trusted for small dipoles and we
take the gluon exchange only into account if one of the dipoles is
smaller than a cut $r_{\rm cut}$. This contribution will only show up
for large $Q^2$, which is not the regime of our main interest. In
addition it is well known that in this regime the much more
sophisticated perturbative approaches do describe the data very well.
Nevertheless we can extend our model to larger values of $Q^2$ by
taking only the leading perturbative contribution into account. We
have to regularize the gluon propagator for large distances, which are
strong suppressed due to $r_{\rm cut}$ anyhow. In order to introduce
as less parameters as possible we use for the cutoff for the gluon
propagator the cut $c$ between the soft- and hard-pomeron. For the
strong coupling we will use a running coupling on the 1-loop level
with $\La_{\rm QCD}=1\fm$, which is frozen in the infra-red to
$\al_s(\infty )$. This procedure results in the following perturbative
contribution to \eq{ddcplus}:
\beqa
\tilde{\ch}_{\rm per}&=&12(N_{\rm C}^2-1)4\pi\left[ \al_s\left(\vec{r}_{1q}-\vec{r}_{2q}\right)\De\left(\vec{r}_{1q}-\vec{r}_{2q}\right)\right. +\al_s\left(\vec{r}_{1\bar{q}}-\vec{r}_{2\bar{q}}\right)\De\left(\vec{r}_{1\bar{q}}-\vec{r}_{2\bar{q}}\right)\nn\\
&&\hspace{2.2cm}-\al_s\left(\vec{r}_{1q}-\vec{r}_{2\bar{q}}\right)\De\left(\vec{r}_{1q}-\vec{r}_{2\bar{q}}\right)-\left.\al_s\left(\vec{r}_{1\bar{q}}-\vec{r}_{2q}\right)\De\left(\vec{r}_{1\bar{q}}-\vec{r}_{2q}\right)\right],
\lbqa{chiper}
where the coupling $\al_s$ is running on the 1-loop level
\[
\al_s(r)= \left\{ {\al_s(\infty )\;|\; r>c}\atop{\al_s(\infty )\frac{\displaystyle\log (1\fm/c)}{\displaystyle\log (1\fm / r)}\;|\;r\le c} \right\}
\]
and $\De$ is the Fourier-transformed of the regularized gluon propagator in the transversal plane
\beq
\De(\vec{x})=\cF_2\left[\frac{1}{\vec{k}^2+\frac{1}{c^2}}\right](\vec{x})=\frac{1}{2\pi}K_0\left(\frac{|\vec{x}|}{c}\right).
\lbq{2dimfourierper}
Taking for small dipoles this perturbative contribution into account \eq{Jtildetwopomeron} becomes
\beqa
&&\int \, d^2b\,e^{-i\vec{\De}_\perp\cdot\vec{b}}\,\frac{1}{8 N_{\rm C}^2(N_{\rm C}^2-1)12^2}\left( \tilde{\ch}^2\,\th(R_1-r_{\rm cut})\,\th(R_2-r_{\rm cut})+\tilde{\ch}_{\rm per}^2\,\tilde\th(R_1,R_2)\right)\label{Jtilde}\\
&&\times\left(f_{\rm h}(R_1,R_2)\left(\frac{W}{20\GeV}\right)^{0.56}+f_{\rm s}(R_1,R_2)\left(\frac{W}{20\GeV}\right)^{2(0.08+0.25{\rm GeV}^{-2}t)}\right)\nn
\end{eqnarray}
where $\tilde\th(R_1,R_2)$ is 1 if $R_1$ or $R_2$ is smaller than $r_{\rm cut}$ and 0 else.

In the next section we present our results for elastic hadron-hadron
scattering, elastic photo- and electroproduction of vectormesons and
the proton structure function $F_2$. We only fitted three parameters:
the cut $c$ between the soft- and hard-pomeron, the strong coupling in
the infra-red $\al_s(\infty)$ and the cut $r_{\rm cut}$ where we cut
the contributions of the MSV. We did not made a real fit to the data
but started with values which are very physical and adjusted only a
little bit. For the cut $r_{\rm cut}$ one expects a value near to 1
GeV and our final value is $r_{\rm cut}=0.16\fm$. For $c$ we obtain
$c=0.35\fm$. The coupling in the infra-red was estimated in our model
to be 0.5 \cite{Rueter:1995,Dosch:1995} and here our final value is
$\al_s(\infty)=0.75$.
\section{Results for the different reactions}
\subsection{Elastic hadron-hadron scattering}
The hadron-hadron profile function is obtained by smearing the dipole-dipole profile function (\eq{Jtilde}) with simple phenomenological wavefunctions for the hadrons. For the hadrons we use a diquark picture as indicated by the good description of elastic hadron-hadron scattering at $\sqrt s = 20 \GeV$ \cite{Dosch:1994} and the suppression of the odderon coupling \cite{Rueter:1996,Rueter:1996III}. We use a simple Gaussian wavefunction and obtain
\beq
J=\int \frac{d^2 r_1}{4\pi}\int \frac{d^2 r_2}{4\pi}\left|\Psi^1 (r_1)\right|^2\left|\Psi^2 (r_2)\right|^2\tilde{J}
\lbq{Jhaha}
with
\[
\Psi^i(r_i)=\frac{\sqrt{2}}{S_i}e^{-\frac{r_i^2}{4S_i^2}}
\]
and $S_i$ being the hadron sizes fitted to the data at $\sqrt s = 20 \GeV$. The wavefunctions are normalized as follows:
\[
\int \frac{d^2 r}{4\pi}\left|\Psi(r)\right|^2=1.
\]
Because of the large sizes of $p$, $\pi$ and $K$ mainly dipoles which are larger than $c$ contribute. Thus we get for $\sqrt s = 20 \GeV$ the same results as in the older publications and describe the data very well. By increasing $s$ our scattering amplitudes now rise like $T\propto s^{1.08}$ and thus all hadronic total cross section rise with the power of the soft-pomeron and do fit the data.
\subsection{Elastic photoproduction of vectormesons}
Considering the process $\ga^* p \rightarrow \mbox{VM } p$ for VM=$\rh,\om,\ph,J/\Ps$ we have the following profile function
\beq
J=\int \frac{d^2 r_{\rm P}}{4\pi}\sum_{f,h_1, h_2}\int \frac{d^2 r_\ga}{4\pi}\int_{z_f}^{1-z_f} dz\Psi^{*\, {\rm VM}}_{f h_1 h_2}(\vec{r}_\ga,z)\Psi^{\ga}_{f h_1 h_2}(\vec{r}_\ga,z)\left|\Psi^{\rm P}(r_{\rm P})\right|^2\tilde{J}.
\lbq{profilepi}
The wavefunctions of the photon and the vectormesons depend on the flavor $f$, the helicity $h_i$ of the (anti)quark and on the momentum fraction $z\;(1-z)$ carried by the quark (antiquark) with respect to the total momentum. Our approach is based on an eikonal approximation \cite{Nachtmann:1991} where the (anti)quarks have to be fast as compared to the fluctuations of the non-trivial QCD vacuum structure. For $z$ near to 0 or 1 the quark or antiquark respectively becomes slow and in the cm-frame the validity of the eikonal approximation thus induces a cut of the $z$ range proportional to $1/W$. For the scale of the cut we take
\[
z_{u,d,s,c}=0.2\GeV/W.
\]
Such an end-point cutting was already discussed in reference
\cite{Dosch:1997II} and introduces an additional energy dependence.
Without this $W$ dependence the scattering amplitude could always be
written as a sum of the two pomerons. Especially for large $Q^2$,
where the end-points of the wavefunction become important, this energy
dependence is important. The wavefunctions are normalized with
\[
\int \frac{d^2 r}{4\pi}dz\sum_{f,h_1, h_2}\left|\Psi_{f h_1 h_2}(r,z)\right|^2=1.
\]
The photon wavefunctions can be computed using light cone perturbation
theory \cite{Bjorken:1971,Lepage:1980}. They depend on the
polarization and virtuality of the photon and are given in the
appendix. These wavefunctions can also be used for small values of
$Q^2$ by introducing running quark masses that depend on the
virtuality and become equal to the constituent masses for $Q^2=0$
\cite{Dosch:1997}. The exact relations are given in the appendix.

For the vectormesons we use the phenomenological wavefunctions derived in reference \cite{Dosch:1997}. For longitudinal polarization we have
\beq
\Psi^{\rm VM}_{f h_1 h_2}(\vec{r},z)=\frac{c_f^{\rm VM}}{\sum_{f'} c_{f'}^{\rm VM}e_{f'}/e}z(1-z)\frac{\de_{h_1,-h_2}}{\sqrt 2}\frac{\sqrt 2 \pi f^{\rm VM}}{\sqrt{N_{\rm C}}}f(z)e^{-\om^2 r^2/2}
\lbq{VMlongwf}
where $c_f^{\rm VM}$ is the Clebsch-Gordan for the flavor $f$ depending on the vectormeson and is given in table \ref{VMpara} in the appendix. With $f^{\rm VM}$ we denote the vectormeson decay constant and $f(z)$ is modeled in a way proposed by Wirbel, Stech and Bauer \cite{Wirbel:1985}:
\beq
f(z)=\cN\sqrt{z(1-z)}e^{-M_{\rm VM}^2(z-1/2)^2/(2\om^2)}.
\lbq{VMf}
The two parameters of these wavefunctions ($\cN, \om$) are fixed by the normalization and the measured meson leptonic decay constant. For small values of $Q^2$ we have to fix these parameters taking the running quark mass into account (see appendix B of reference \cite{Dosch:1997}). Our results are given in table \ref{VMpara} in the appendix together with the wavefunctions for transversal polarized vectormesons.

With these formulae we calculate the scattering amplitude of the vectormeson production. In this paper we concentrate on total cross sections leaving for example the slope for following publications. In \fig{photo} we show the result for the total elastic cross section of photoproduction of $\rh ,\om$ and $\ph$ as compared to experimental values.
\befi{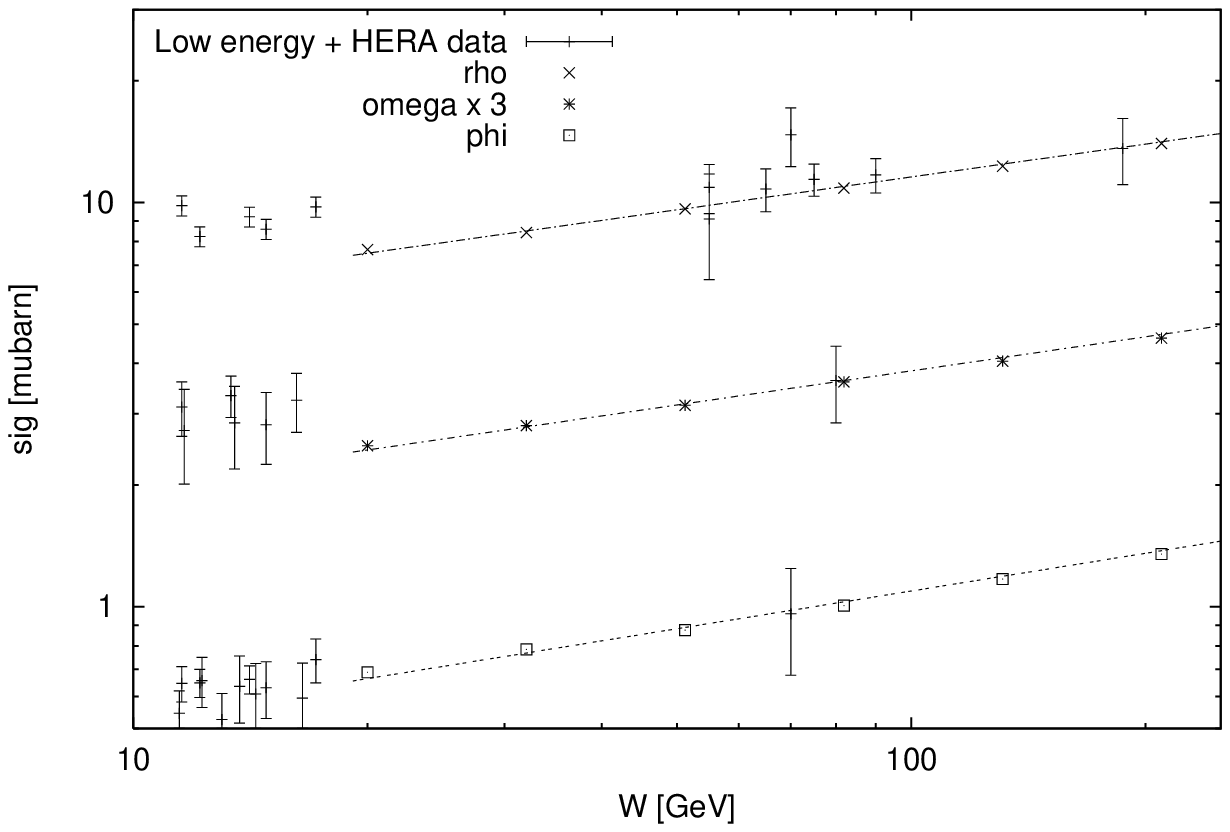}{10cm}
\unitlength1cm
\lbcap{14cm}{The total elastic cross section of photoproduction of $\rh ,\om$ and $\ph$ as compared to low-energy data \protect\cite{Egloff:1979,Busenitz:1989} and HERA data \protect\cite{Aid:1996,Derrick:1995,Breitweg:1997,Derrick:1996,Derrick:1996II}. The $\om$ data are scaled with a factor 3. The lines represent our exponential fit described in the text.}{photo}

To obtain the effective energy dependence we used a simple exponential fit
\beq
\si^{\rm tot}=a\left(\frac{W}{20\GeV}\right)^b
\lbq{JPsifit1}
with the result
\[
\begin{tabular}{|c|c|c|c|}
\hline
&$\rh$&$\om$&$\ph$\\
\hline\hline
$a[\mu {\rm b}]$&7.49&0.814&0.664\\
\hline
$b$&0.27&0.28&0.31\\
\hline
\end{tabular}
\]
The power is near to the pure soft-pomeron which shows that for these
processes mainly large dipoles, larger than the cut $c$, contribute
and the hard-pomeron is negligible. In our results only the
pomeron-part is included and not the contribution from
Regge-trajectories. This explains why our results would underestimate
the data for $W\le 20 \GeV$, especially for the $\rh$ and $\om$. We
also observe that $b$ increases slightly by going from $\rh$ to $\ph$
which is due to the increasing mass and thus the smaller size of the
meson.

Now we come to the more interesting case of $J/\Ps$ photoproduction. Our result is shown in \fig{photoJPsi}.
\befi{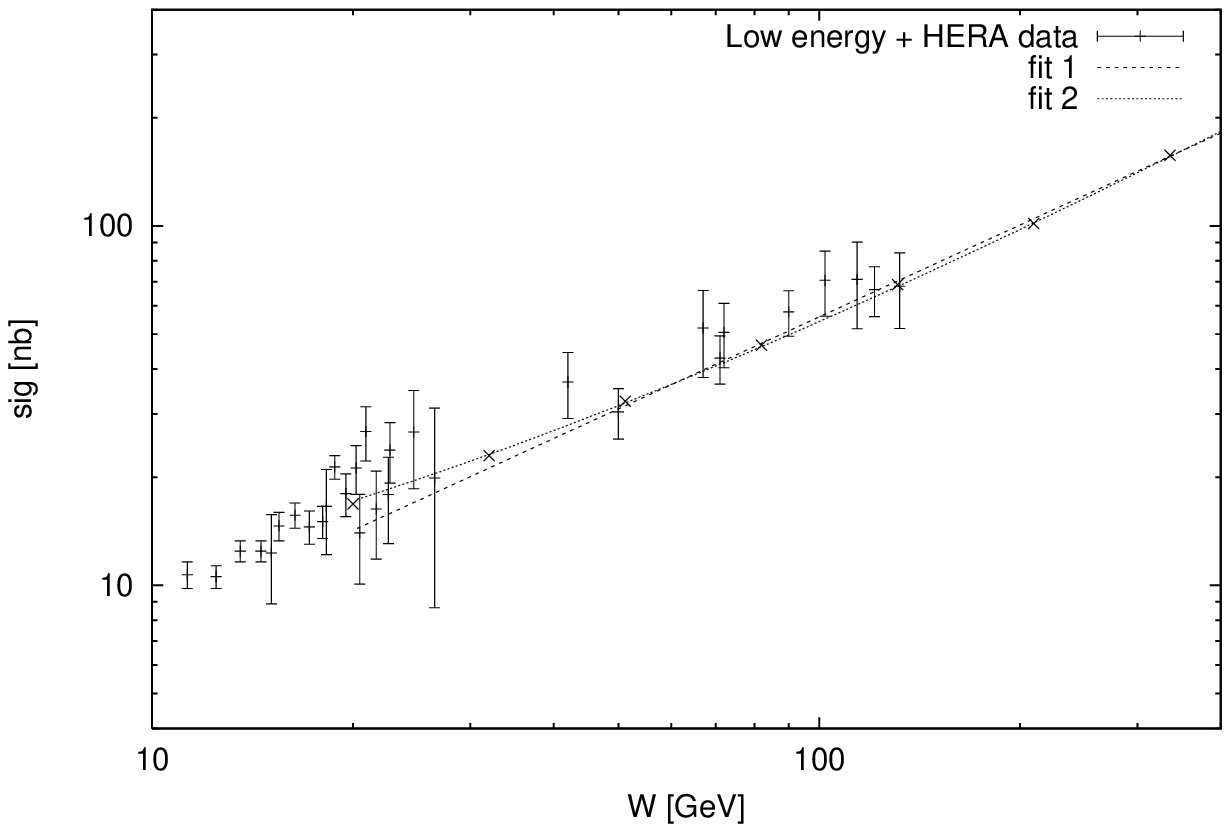}{10cm}
\unitlength1cm
\lbcap{14cm}{The total elastic cross section of photoproduction of $J/\Ps$ (crosses) as compared to low-energy data \protect\cite{Binkley:1982,Frabetti:1993} and HERA data \protect\cite{Derrick:1995II,Aid:1996II,Breitweg:1997II}. Fit 1 represents again our exponential fit (\eq{JPsifit1}) and fit 2 is a fit with two powers (\eq{JPsifit2}). There are also preliminary H1 data for larger $W$ (see for example reference \protect\cite{H1:1997}) which confirm the concave behavior of our result.}{photoJPsi}

The exponential fit to our results with large $W$ gives
\[
\begin{tabular}{|c|c|}
\hline
&$J/\Ps$\\
\hline\hline
$a[{\rm nb}]$&14.22\\
\hline
$b$&0.85\\
\hline
\end{tabular}
\]
but the data have the tendency to grow with a higher and higher power for large $W$. The large value of $b=0.85$ shows that for the $J/\Ps$ small dipoles are important. Our data can be well described over the whole $W$ range by a fit with two powers, where we fix one to be 0.22 which is approximately the behavior of the pure soft-pomeron in this $W$ range:
\beq
\si^{\rm tot}=a\left(\frac{W}{20\GeV}\right)^{0.22}+b\left(\frac{W}{20\GeV}\right)^c
\lbq{JPsifit2}
and we obtain
\[
\begin{tabular}{|c|c|}
\hline
&$J/\Ps$\\
\hline\hline
$a[{\rm nb}]$&8.88\\
\hline
$b[{\rm nb}]$&8.31\\
\hline
$c$&1.00\\
\hline
\end{tabular}
\]
which shows that to photoproduction of $J/\Ps$ the hard- and soft-pomeron contribute with similar size.

Because for $J/\Psi$ the hard-pomeron is important it is also interesting to investigate how much the total result is due to the contributions calculated within the MSV (dipoles larger than $r_{\rm cut}$) or due to the two gluon exchange $\tilde \ch_{\rm per}$. Therefor we show in \fig{photoJPsinonper} the contribution of $\tilde \ch$, the nonperturbative part calculated in the MSV, as compared to the total cross section.
\befi{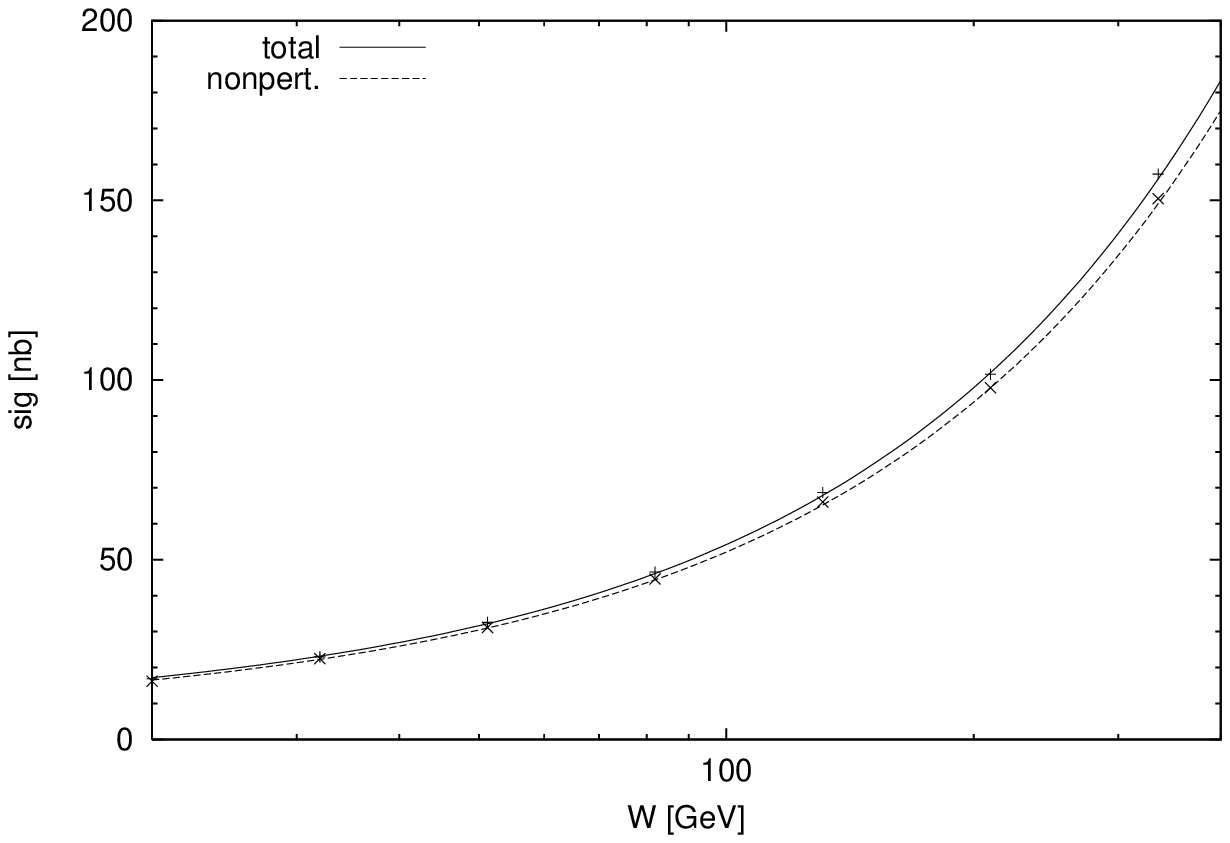}{10cm}
\unitlength1cm
\lbcap{14cm}{The total elastic cross section of photoproduction of $J/\Ps$ with all contributions and only the nonperturbative contribution.}{photoJPsinonper}

From \fig{photoJPsinonper} we conclude that the dipoles for $J/\Ps$
photoproduction are larger then the cut $r_{\rm cut}$. The strong rise
with $W$ is due to dipole sizes between $r_{\rm cut}$ and $c$ which
are treated with the MSV but are increased with the power of the
hard-pomeron.
\subsection{Elastic electroproduction of vectormesons}
We consider in this subsection the electroproduction of vectormesons.
Our main interest is to show how our model can describe the rising
effective pomeron power with rising virtuality $Q^2$. Therefor we
concentrate on the $\rh$ and $J/\Ps$ meson and only investigate the
total cross sections. We leave the study of the $\om$ and $\ph$ and
the discussion of the differential cross section or the different
behavior of the longitudinal and transversal contribution for
following publications.

In \fig{electroRhoQalle} we show our result for the electroproduction of the $\rh$ meson for different values of $Q^2$ as a function of $W$.
\befi{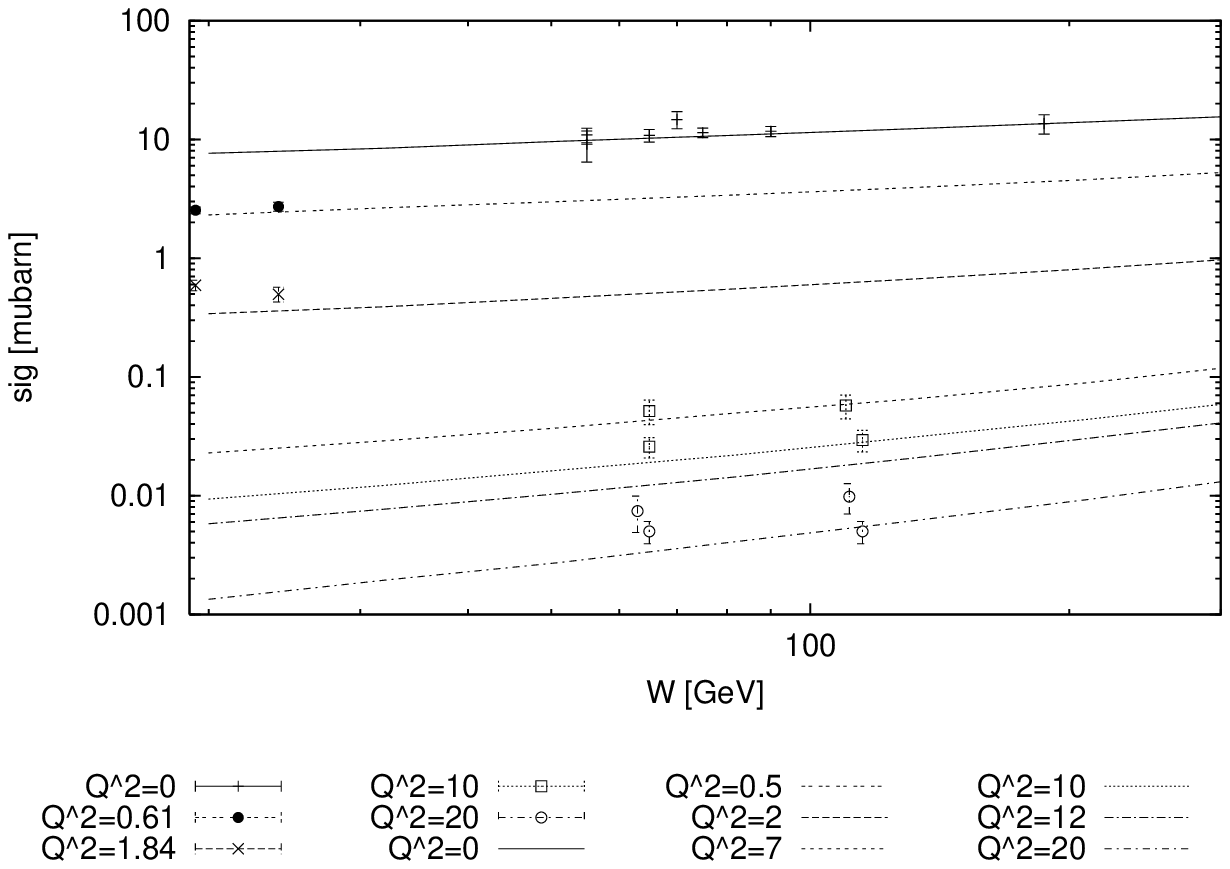}{10cm}
\lbcap{14cm}{The total elastic cross section of photo- and electroproduction of $\rh$. The experimental electroproduction data are from H1 \protect\cite{Aid:1996IIII} for $Q^2=10{\;\rm GeV}^2$ and $20{\;\rm GeV}^2$. The ZEUS data \protect\cite{Derrick:1995IIII} are scaled to the same $Q^2$ values and the low-energy data are from reference \protect\cite{Adams:1997}. There are new preliminary HERA data (see for example \protect\cite{ZEUS:1997,H1:1997II}) and our results are shown at some $Q^2$ values of these analysis.}{electroRhoQalle}

We obtain a quite good description of the data taking also the
preliminary results \cite{ZEUS:1997,H1:1997II} into account. Our
results are always below the ZEUS data which are larger then the H1
data for the same $Q^2$. The ZEUS data have a large overall error due
to normalization uncertainties which is not included in the figure.
For the whole $W$ range our results can be well described by the two
power fit (\eq{JPsifit2}) with the result
\[
\begin{tabular}{|c|c|c|c|c|c|c|}
\hline
$Q^2[\GeV^2]$&0.5&2&7&10&12&20\\
\hline\hline
$a[{\rm nb}]$&2264&287&17.2&5.65&3.26&0.654\\
\hline
$b[{\rm nb}]$&76.2&51.4&6.40&3.83&2.57&0.740\\
\hline
$c$&1.00&0.80&0.96&0.94&0.96&1.03\\
\hline
\end{tabular}
\]
The simple exponential fit (\eq{JPsifit1}) describes our result very good for large $W>80 \GeV$ and we obtain
\[
\begin{tabular}{|c|c|c|c|c|c|c|}
\hline
$Q^2[\GeV^2]$&0.5&2&7&10&12&20\\
\hline\hline
$a[{\rm nb}]$&2064&291&17.5&7.25&4.43&1.08\\
\hline
$b$&0.34&0.44&0.71&0.77&0.82&0.92\\
\hline
\end{tabular}
\]
The rising of $b$ with $Q^2$ shows that by increasing the virtuality one probes smaller and smaller dipoles which are coupled to the hard-pomeron. This rise of the effective pomeron power with $Q^2$ is in agreement with the experiment.

In \fig{electroJPsi} we show our result for the electroproduction of the $J/\Ps$ for different values of $Q^2$ as a function of $W$.
\befi{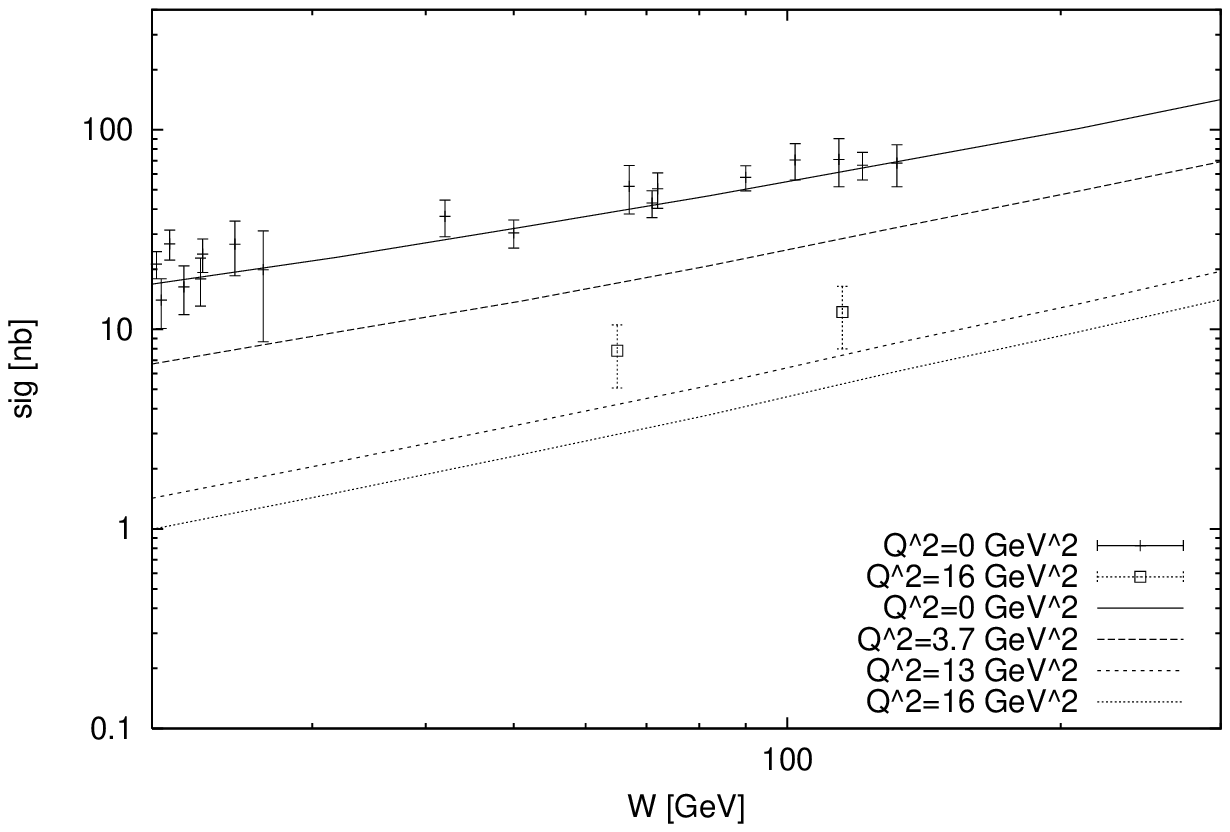}{10cm}
\lbcap{14cm}{The total elastic cross section of photo- and electroproduction of $J/\Ps$. The experimental electroproduction data are from H1 \protect\cite{Aid:1996IIII} for $Q^2=16{\;\rm GeV}^2$. There are new preliminary HERA data (see for example \protect\cite{ZEUS:1997}) and our results are shown at some $Q^2$ values of these analysis. For $Q^2=3.7{\;\rm GeV}^2$ and $Q^2=13{\;\rm GeV}^2$ our result agrees well with these preliminary data.}{electroJPsi}

As can be seen from \fig{electroJPsi} we underestimate the H1 data for
$Q^2=16\GeV^2$ but comparison with the new preliminary HERA data (see
for example \protect\cite{ZEUS:1997}) is more satisfactory. For these
large values of $Q^2$ our results can only be trusted for larger
values of $W$. Thus we did not include values with $W<20\GeV$ in the
plot where more experimental data exist. Our results can be well
described by the simple exponential fit (\eq{JPsifit1}) with the
result
\[
\begin{tabular}{|c|c|c|c|}
\hline
$Q^2[\GeV^2]$&3.7&13&16\\
\hline\hline
$a[{\rm nb}]$&5.86&1.27&0.903\\
\hline
$b$&0.91&1.01&1.02\\
\hline
\end{tabular}
\]
Again $b$ is rising by increasing $Q^2$ but for the $J/\Ps$ already for photoproduction the dipoles are quite small and thus the effect is here not as dramatic as for the $\rh$ meson.
\subsection{The proton structure function $F_2(x,Q^2)$ and the total cross section of $\ga$-$p$ scattering}
Now we come to the proton structure function $F_2(x,Q^2)$, that is the total cross section of $\ga^*$-$p$ scattering. The profile function is given by
\beq
J=\int \frac{d^2 r_{\rm P}}{4\pi}\sum_{f,h_1, h_2}\int \frac{d^2 r_\ga}{4\pi}\int_{z_f}^{1-z_f} dz\left|\Psi^{\ga}_{f h_1 h_2}(\vec{r}_\ga,z)\right|^2\left|\Psi^{\rm P}(r_{\rm P})\right|^2\tilde{J}.
\lbq{profileF2}
Using the different photon polarizations we obtain with \eq{sigtotundslope} and \eq{dsigdt} the total cross sections $\si_{\rm L}$, $\si_{\rm T}$ and the proton structure function $F_2$ and $F_{\rm L}$ can be calculated
\beqa
F_2&=&\frac{1}{4\pi^2\al_{\rm em}}\frac{Q^4(1-x)}{Q^2+4m_{\rm P}^2 x^2}\left(\si_{\rm L}+\si_{\rm T}\right)\nn\\
F_{\rm L}&=&\frac{1}{4\pi^2\al_{\rm em}}\frac{Q^4(1-x)}{Q^2+4m_{\rm P}^2 x^2}\si_{\rm L}.
\lbqa{F2Fl}
The energy $W$ can be expressed by $W^2=Q^2/x-Q^2+m_{\rm P}^2$. In our approach we are limited to large energies, $W>20 \GeV$, because we only take the pomeron and not the Regge contributions into account. Also the value of $x$ is limited in our approach. It has to be small enough because we can only describe soft interactions with $W$ being the largest scale. In the plots our result is always shown for $x<0.05$. In \fig{F21} and \fig{F22}, presented at the end of this section, we compare our result for $F_2$ with the experimental data for $0.11 \GeV^2\le Q^2 \le 5000\GeV^2$ and $x$-values as described above.

The figures show that our model describes all the data in the restricted $W$ and $x$ range very well. We want to remind, that only for $Q^2\ge 35 \GeV^2$ the contributions coming from very small dipoles, smaller than $r_{\rm cut}$, are important in our approach. But this is the regime where the more sophisticated perturbative approaches work very well. In our approach we will concentrate mainly on the behavior for $0\le Q^2\le 35\GeV^2$ where we observe the transition from the soft- to the hard-pomeron. To obtain this effective energy dependence of the structure function for different scales we calculate the effective pomeron power $\la_{\rm eff}$ by fitting our results for $10^{-4}\le x\le 10^{-2}$ with
\beq
F_2(x,Q^2)= a\frac{Q^4(1-x)}{Q^2+4m_{\rm P}^2 x^2}W^{2\la_{\rm eff}}=a\frac{Q^4(1-x)}{Q^2+4m_{\rm P}^2 x^2}\left(\frac{Q^2}{x}-Q^2+m_{\rm P}^2\right)^{\la_{\rm eff}}
\lbq{lambdaefffit}
for fixed $Q^2$. The result is shown in \fig{lambdaeff}.
\befi{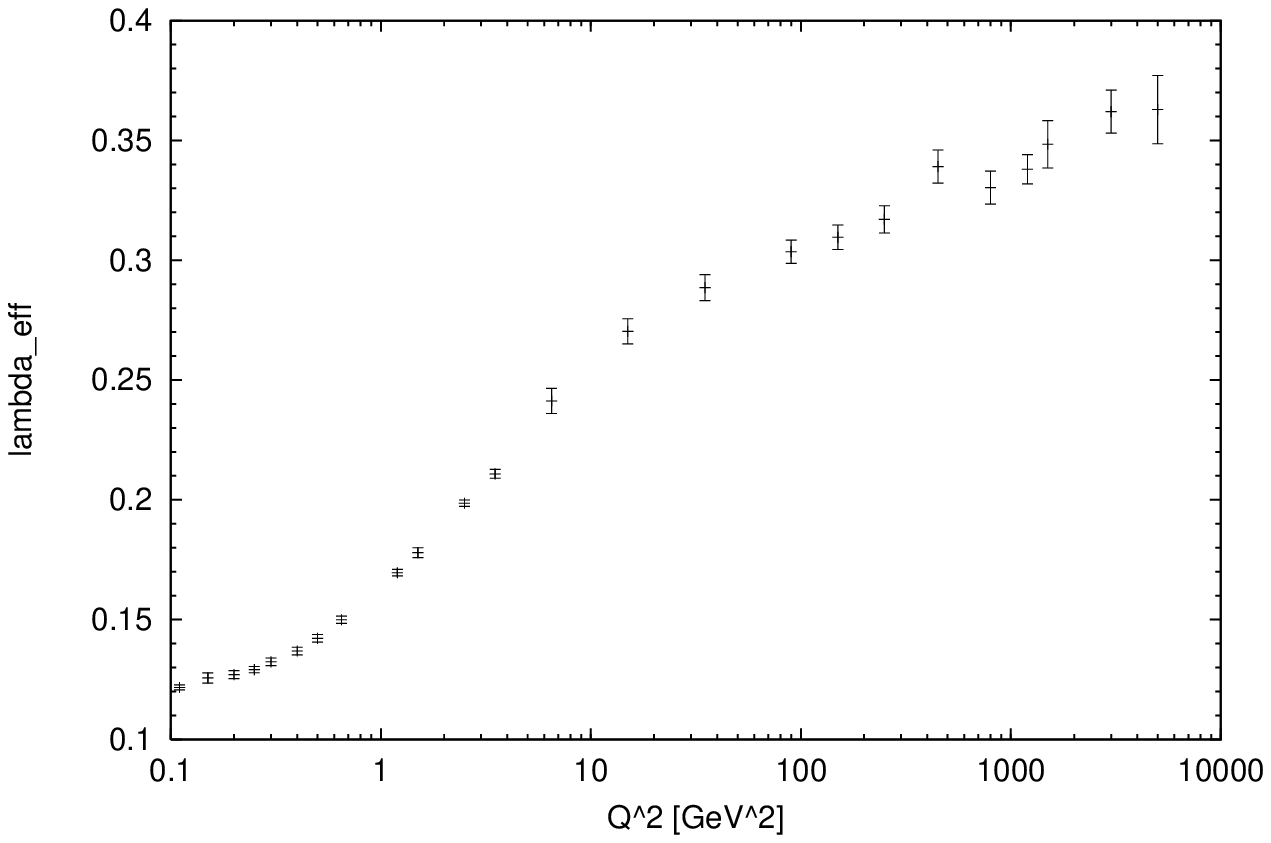}{8.5cm}
\lbcap{14cm}{The effective pomeron power $\la_{\rm eff}$ as defined in \eq{lambdaefffit} as a function of $Q^2$. The error bars are due to the numerical error of our results for $F_2$.}{lambdaeff}

Our result is in very good agreement with published experimental data \cite{Aid:1996III,Adloff:1997} and preliminary ZEUS data. The transition from the soft-pomeron at low $Q^2$ to the higher power at large virtuality can be observed. This transition takes place between 1 and 10 $\GeV^2$. This result shows also the importance of the energy dependence of the photon wavefunction. Without this end-point cutting the maximal $\la_{\rm eff}$ would be 0.28 but this is enhanced to 0.36 which is in good agreement with experiment. But we want to remind that the behavior for asymptotic large $W$, that is asymptotic small $x$ is always given by the hard-pomeron that is 0.28. This effective $W$ dependence in the considered $x$ range is also the reason why the two pomeron fit, $F_2=a\,x^{-0.08}+b\,x^{-0.28}$, would lead to unphysical, negative $a$'s for large $Q^2>20\GeV^2$. Nevertheless for $Q^2\le 20 \GeV^2$ this fit does describe our result well and the result for $a$ and $b$ as function of $Q^2$ is shown in \fig{Weff}.
\befi{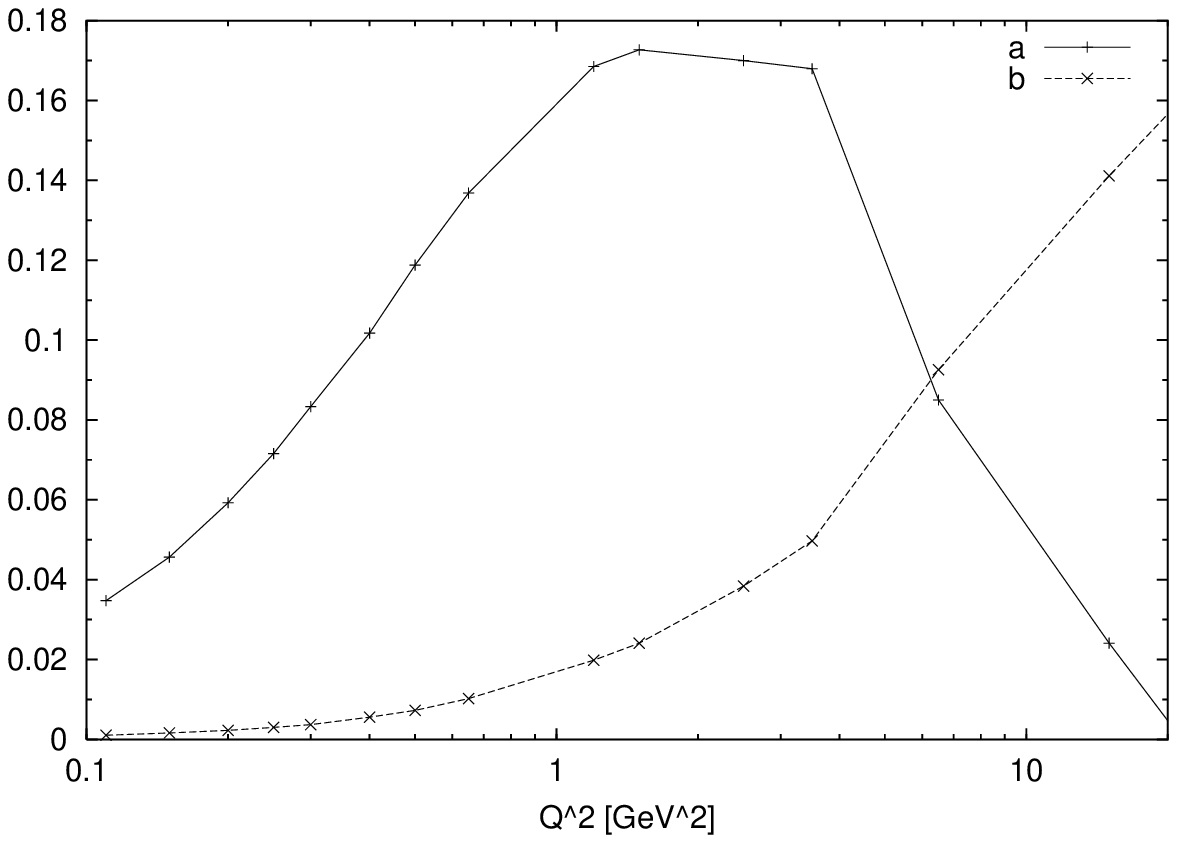}{8.5cm}
\lbcap{14cm}{The parameter $a$ and $b$ of the fit $F_2=a\,x^{-0.08}+b\,x^{-0.28}$ as a function of $Q^2$.}{Weff}

Finally we discuss the dependence of $F_2$ for fixed $x$ on $Q^2$. To do so we calculate the so called Q-slope
\beq
\frac{\pa F_2(x,Q^2)}{\pa \ln Q^2}
\lbq{Qslopedef}
which would be independent of $Q^2$ if the data could be fitted by
\beq
F_2(x={\rm fixed},Q^2)=a+b\ln (Q^2).
\lbq{Qslopefit1}
It turns out that this fit does not work well. So one has to specify the $Q^2$ value at which one calculates the Q-slope for given $x$. Following the analysis of the experimental data in reference \cite{Caldwell:1997} we use $Q^2$ values given like in reference \cite{Desgrolard:1998} by
\beq
Q^2_x=3.1\cdot 10^3\,x^{0.82}.
\lbq{Qvonx}
In \fig{Qslope} we show our result for the Q-slope as a function of $x$.
\befi{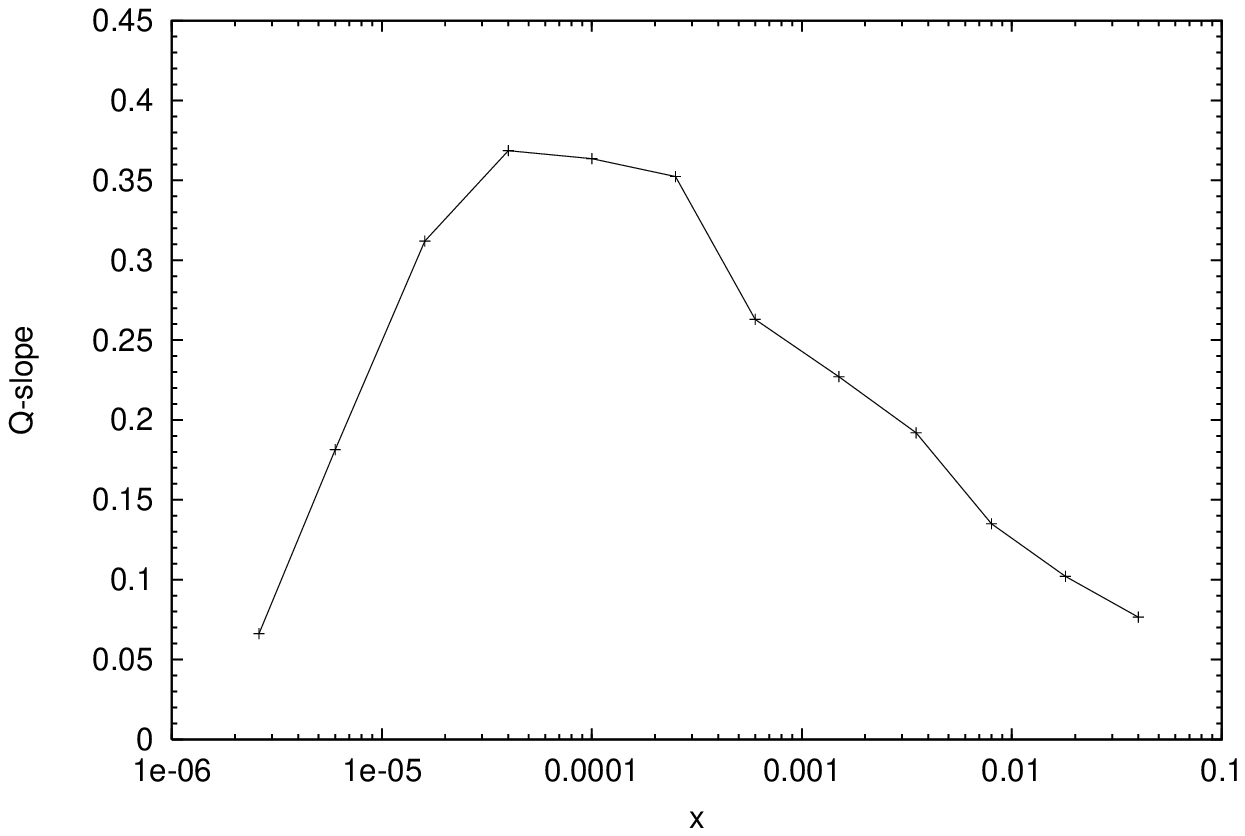}{8.5cm}
\lbcap{14cm}{The Q-slope as a function of $x$ calculated at values for $Q^2$ given by \eq{Qvonx}.}{Qslope}

To compare with the experimental data we present in \fig{Qslopeexp} a plot of an analysis of the HERA data done by Abramowicz and Levy \cite{Abramowicz:1997}. From \fig{Qslope} and \fig{Qslopeexp} we conclude that we describe the $Q^2$ dependence very well except for the data at large $x$. Especially we find that the Q-slope falls for small $x$ after it reaches a maximum at $x\approx 1\cdot 10^{-4}$.
\befi{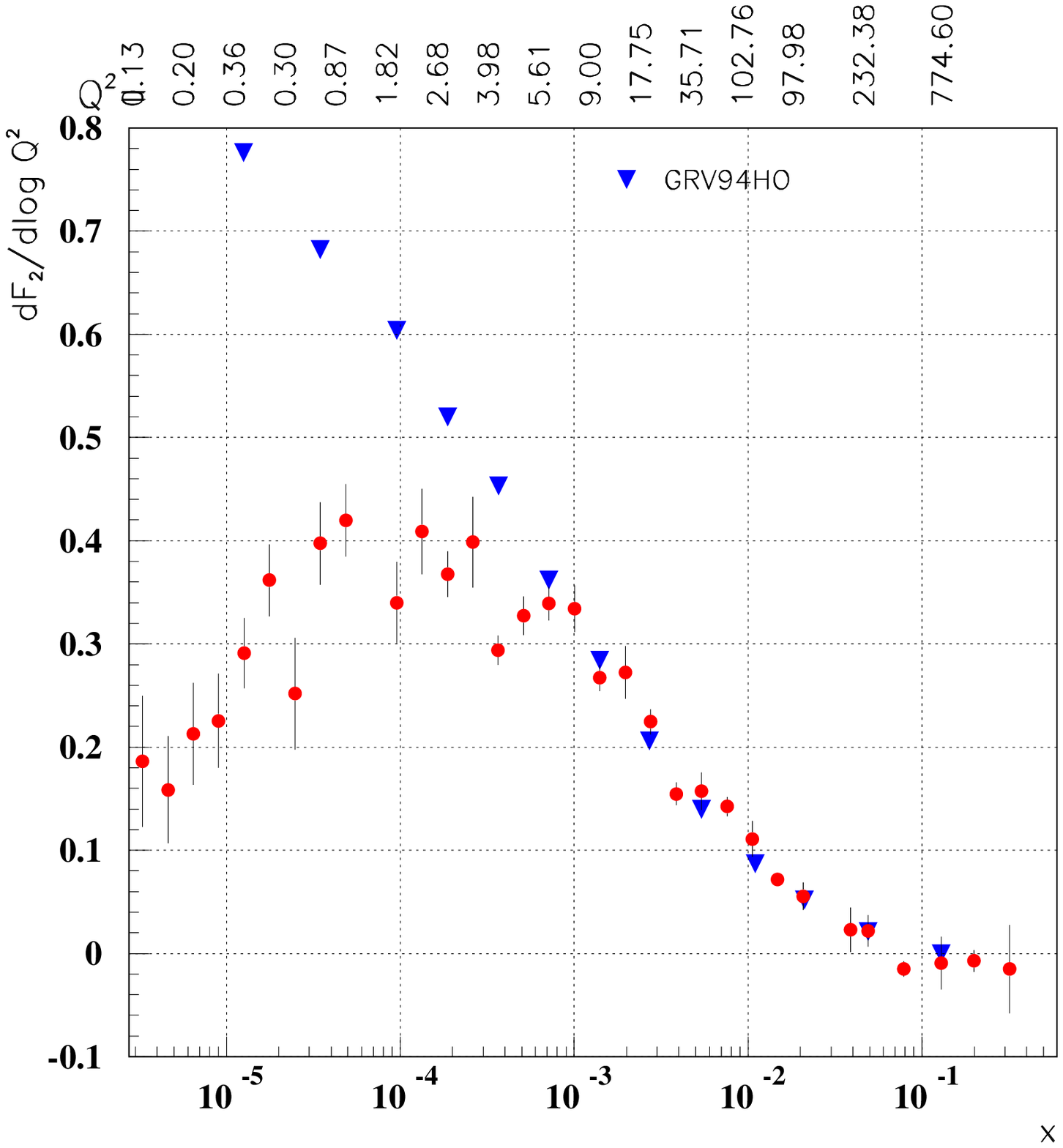}{8.5cm}
\lbcap{14cm}{The Q-slope as a result of an analysis of HERA data \protect\cite{Caldwell:1997} as a function of $x$. In this figure, which is taken from reference \protect\cite{Abramowicz:1997}, also the result of the GRV94 parameterization is shown which can not describe the low-$x$ behavior.}{Qslopeexp}

We have also calculated the total cross section of $\ga^*$-$p$
scattering where we can include the photoproduction. To compare with
experimental data we show in \fig{sigtotWfest} the total cross section
as a function of $Q^2$ for fixed $W$.

From \fig{sigtotWfest} we conclude that we obtain the right $Q^2$
dependence for all values of $W$. Especially the transition at
$Q^2\approx 0.4\GeV^2$ is clearly predicted. The photoproduction
values for large $W$ are also in very good agreement whereas we
underestimate the value at $W=20\GeV$. The reason for this is that the
Regge contributions are important and not included in our model as
already pointed out in reference \cite{Dosch:1997II}.

After discussing the results for $F_2$ we investigate in the following different contributions to the structure function. We will concentrate on the charm contribution $F_2^c$ and on the ratio of the longitudinal to the transversal cross section
\beq
R_{\rm L/T}=\frac{\si_{\rm L}}{\si_{\rm T}}.
\lbq{Rtranslong}
In \fig{F2c} we compare our results for $F_2^c$ at fixed $Q^2$ as a function of $x$ with experiment.

As can be seen from \fig{F2c} our charm contribution is in good agreement with the data for not too large $x$.

Finally we show in \fig{Rlongtrans} the ratio of the longitudinal to the transversal cross section for fixed $W$ as a function of $Q^2$.

\befi{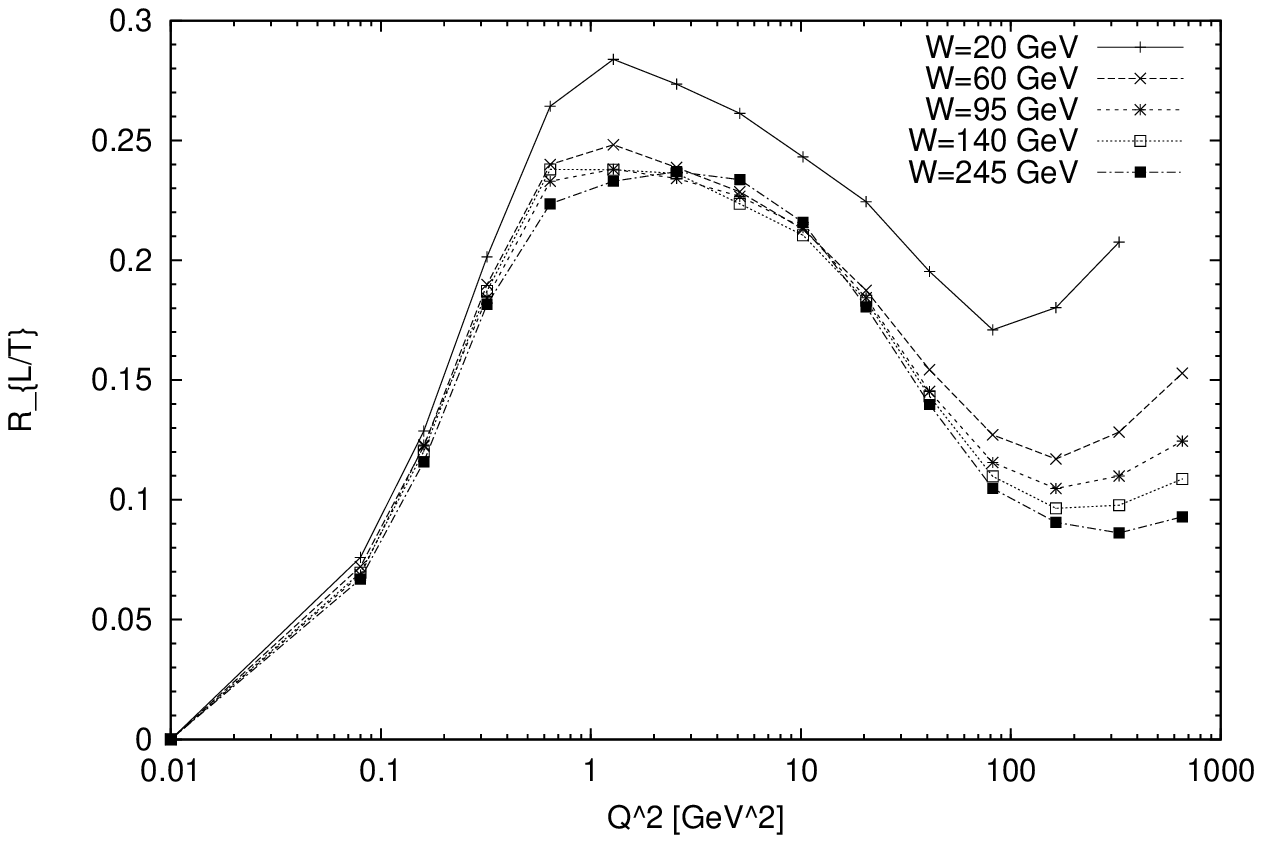}{8.5cm}
\lbcap{14cm}{The ratio of the longitudinal to the transversal cross section $R_{\rm L/T}$ for fixed $W$ as a function of $Q^2$. At $Q^2=0.01{\rm\;GeV}^2$ we put the photoproduction point.}{Rlongtrans}

As can be seen from \fig{Rlongtrans} $R_{\rm L/T}$ is rather independent of $W$ for large $W$. At $Q^2\approx 1\GeV^2$ it reaches a maximum where the longitudinal part is about 25\%. For large $Q^2$, that is large $x$ we observe a rise of $R_{\rm L/T}$ which is maybe just an artifact of our bad description at this kinematic regime. But nevertheless our results indicate that $R_{\rm L/T}$ flattens off at $Q^2 \approx 100 \GeV^2$. Our result for $R_{\rm L/T}$ is quite similar to the one obtained in reference \cite{Martin:1998}.
\begin{figure}[ht]
\leavevmode
\begin{minipage}{5.2cm}
\includegraphics[width=5.2cm]{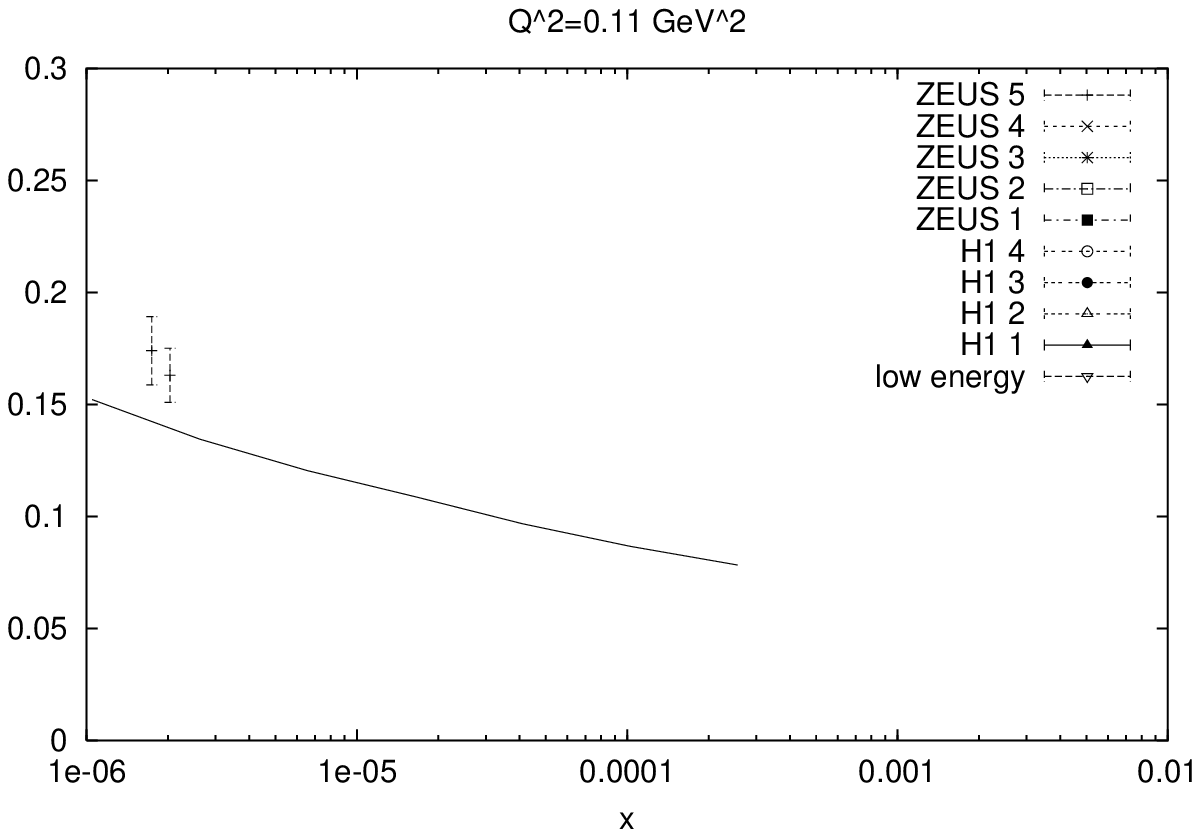}
\end{minipage}
\begin{minipage}{5.2cm}
\includegraphics[width=5.2cm]{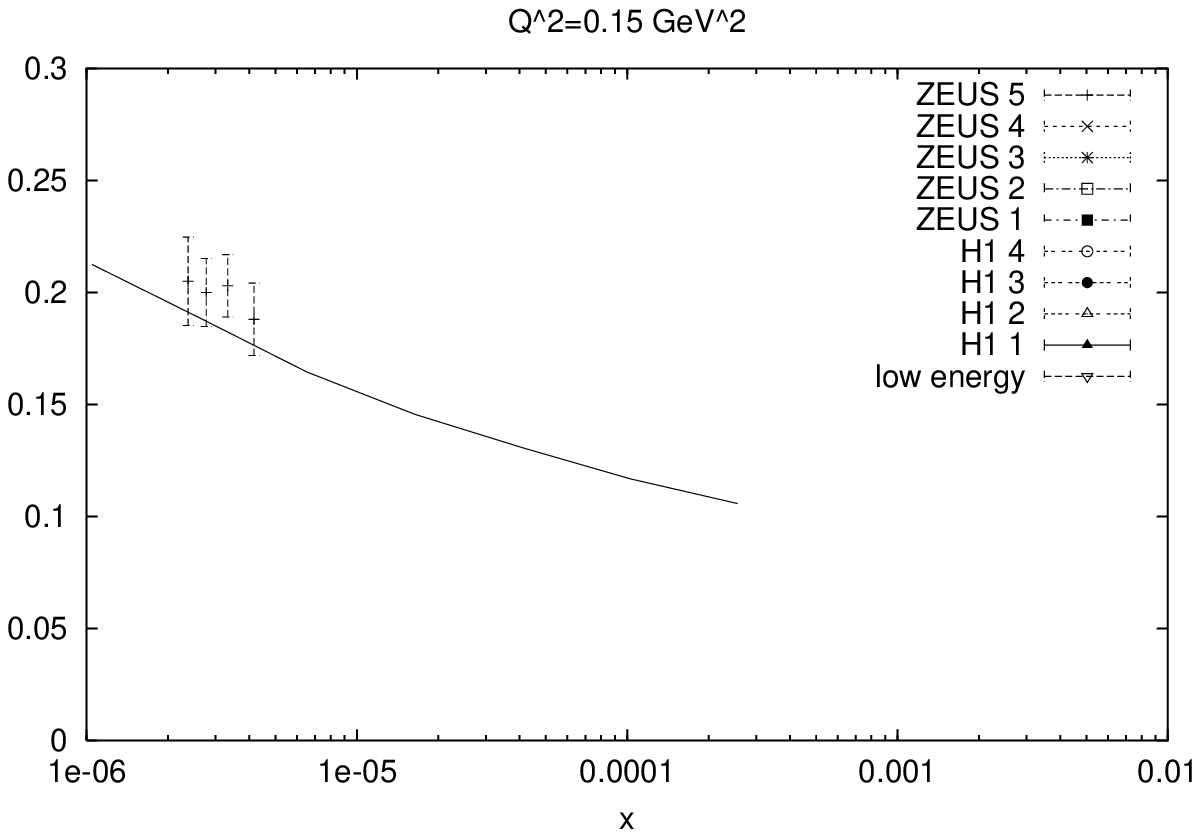}
\end{minipage}
\begin{minipage}{5.2cm}
\includegraphics[width=5.2cm]{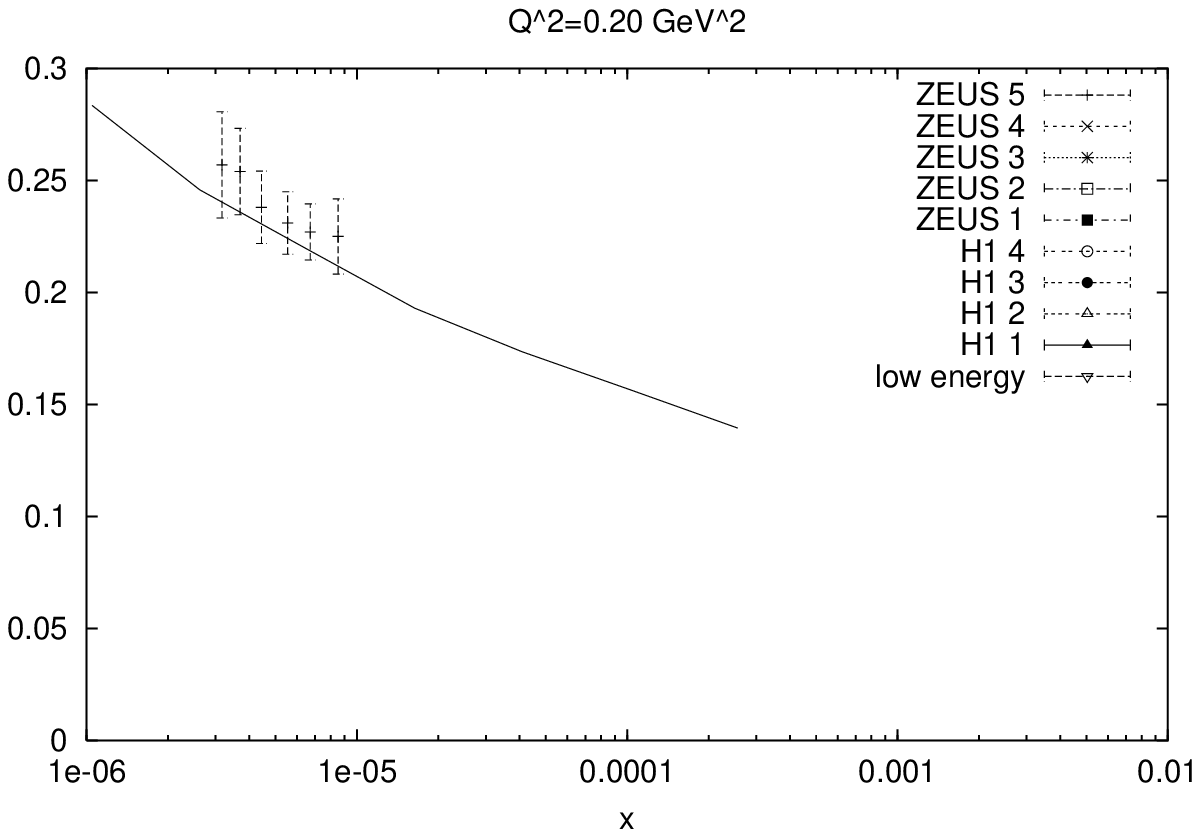}
\end{minipage}
\begin{minipage}{5.2cm}
\includegraphics[width=5.2cm]{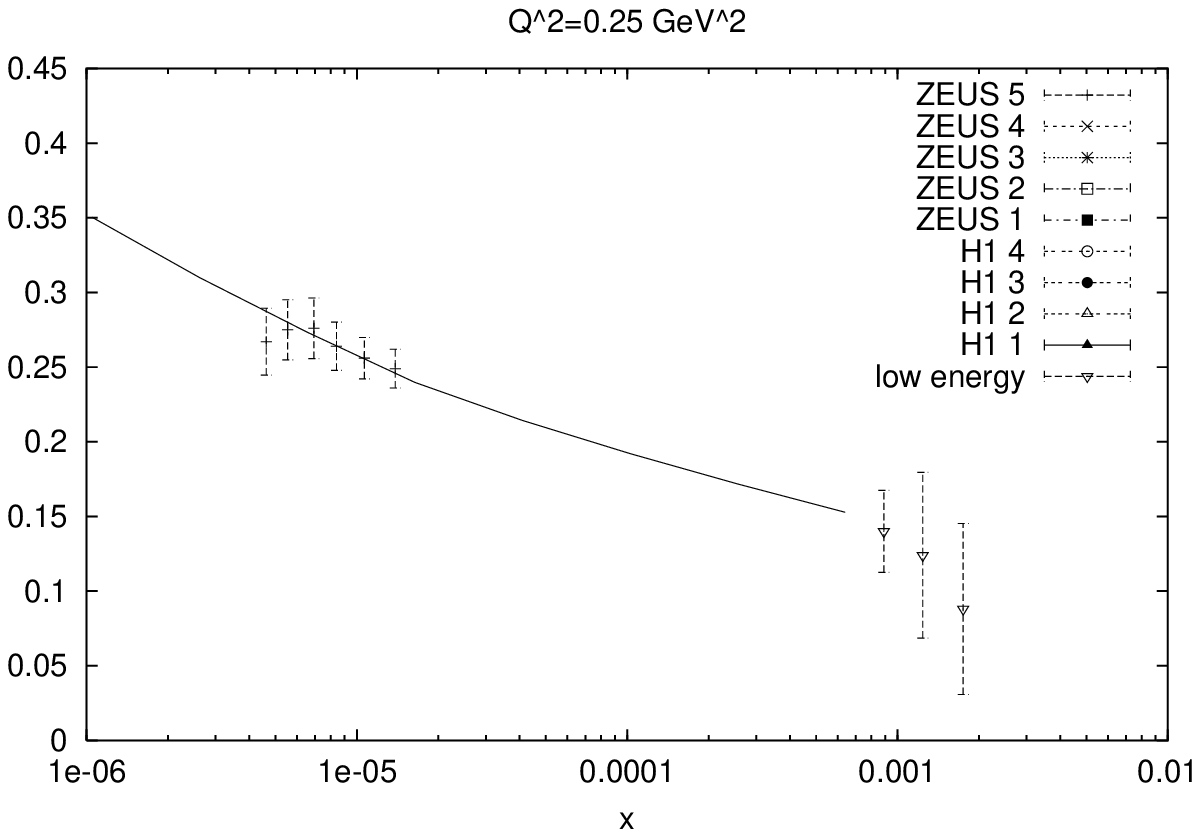}
\end{minipage}
\begin{minipage}{5.2cm}
\includegraphics[width=5.2cm]{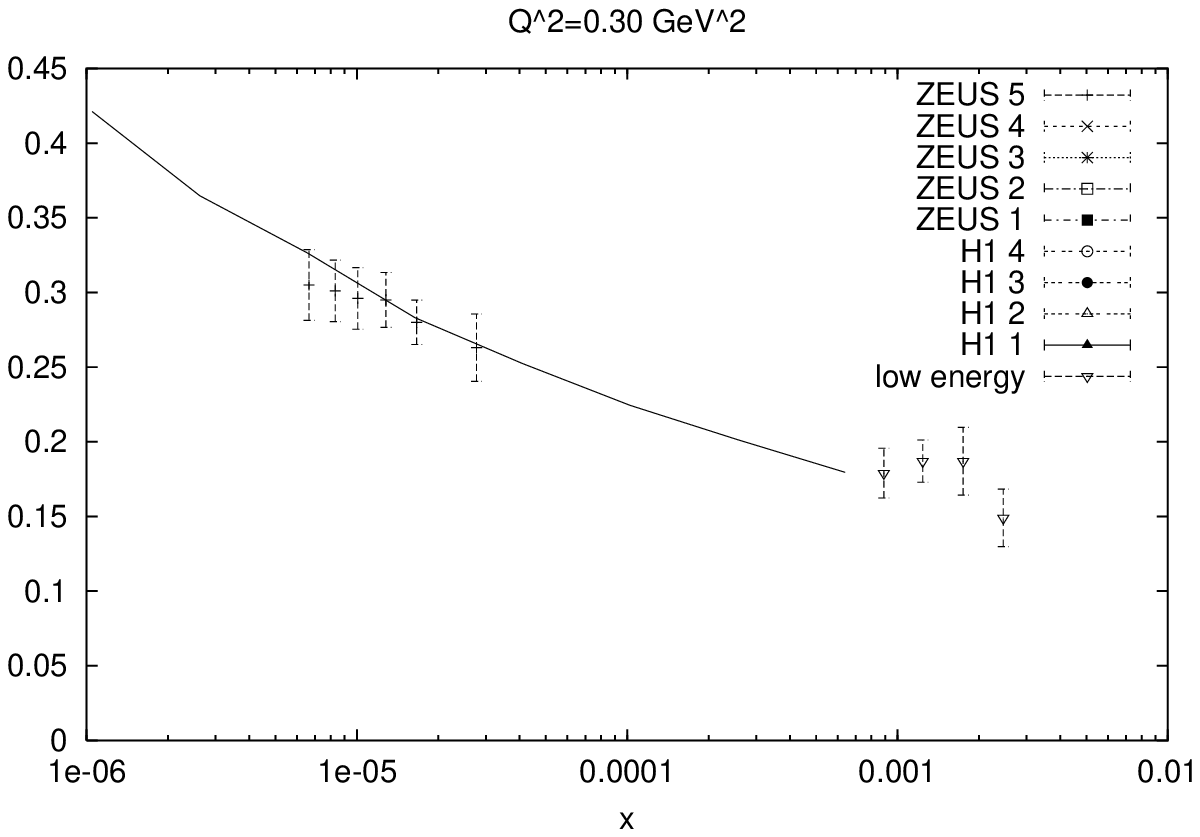}
\end{minipage}
\begin{minipage}{5.2cm}
\includegraphics[width=5.2cm]{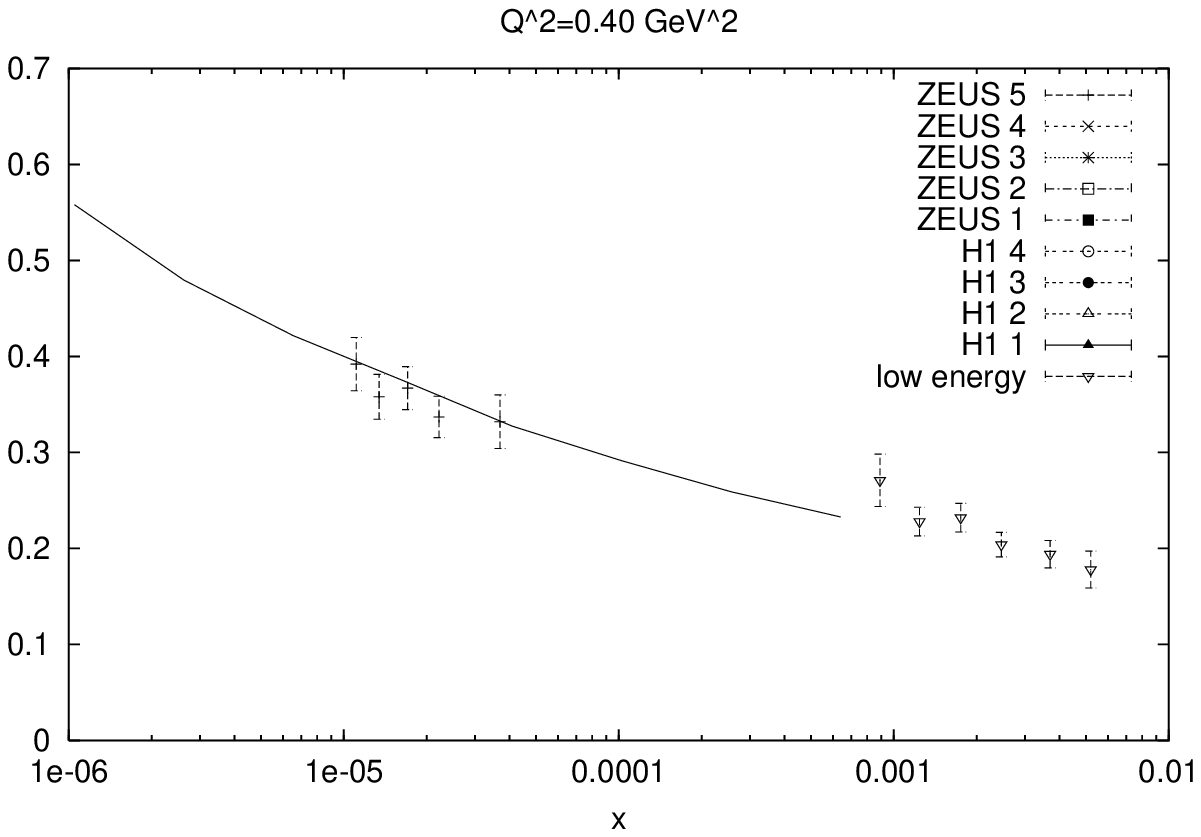}
\end{minipage}
\begin{minipage}{5.2cm}
\includegraphics[width=5.2cm]{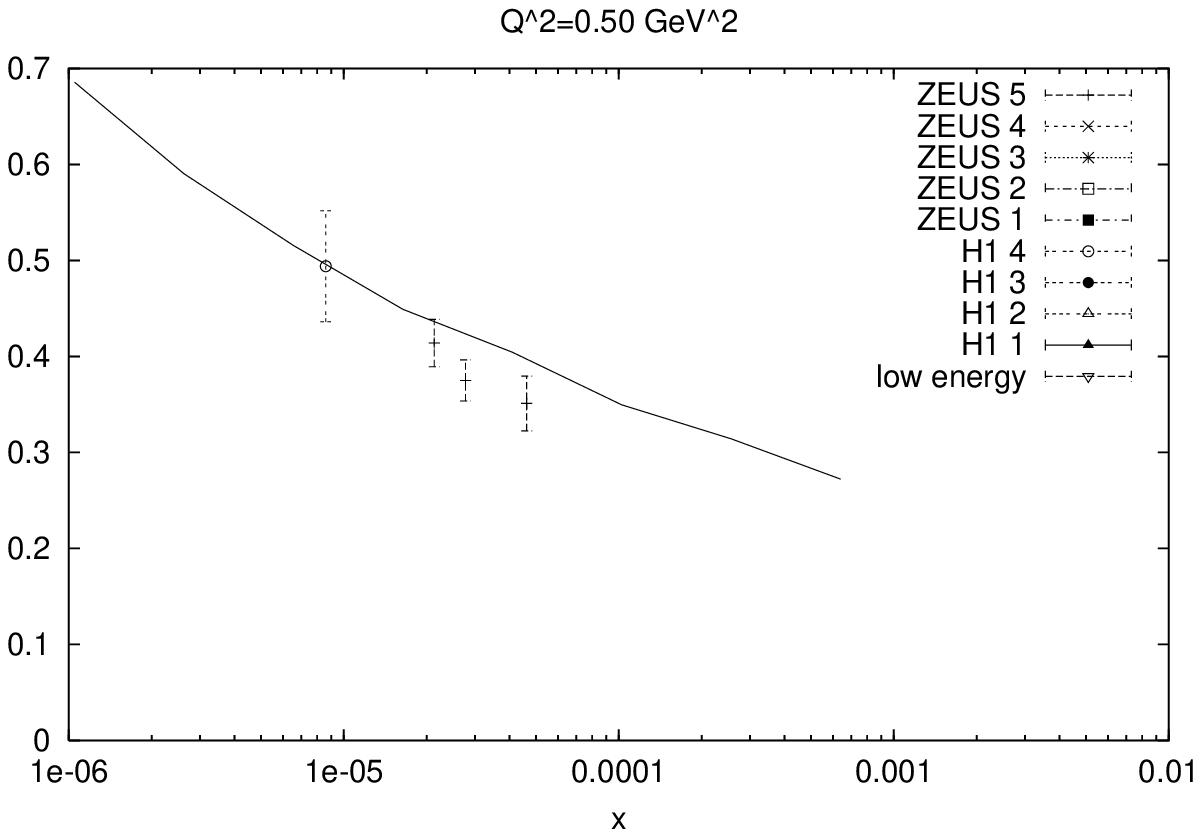}
\end{minipage}
\begin{minipage}{5.2cm}
\includegraphics[width=5.2cm]{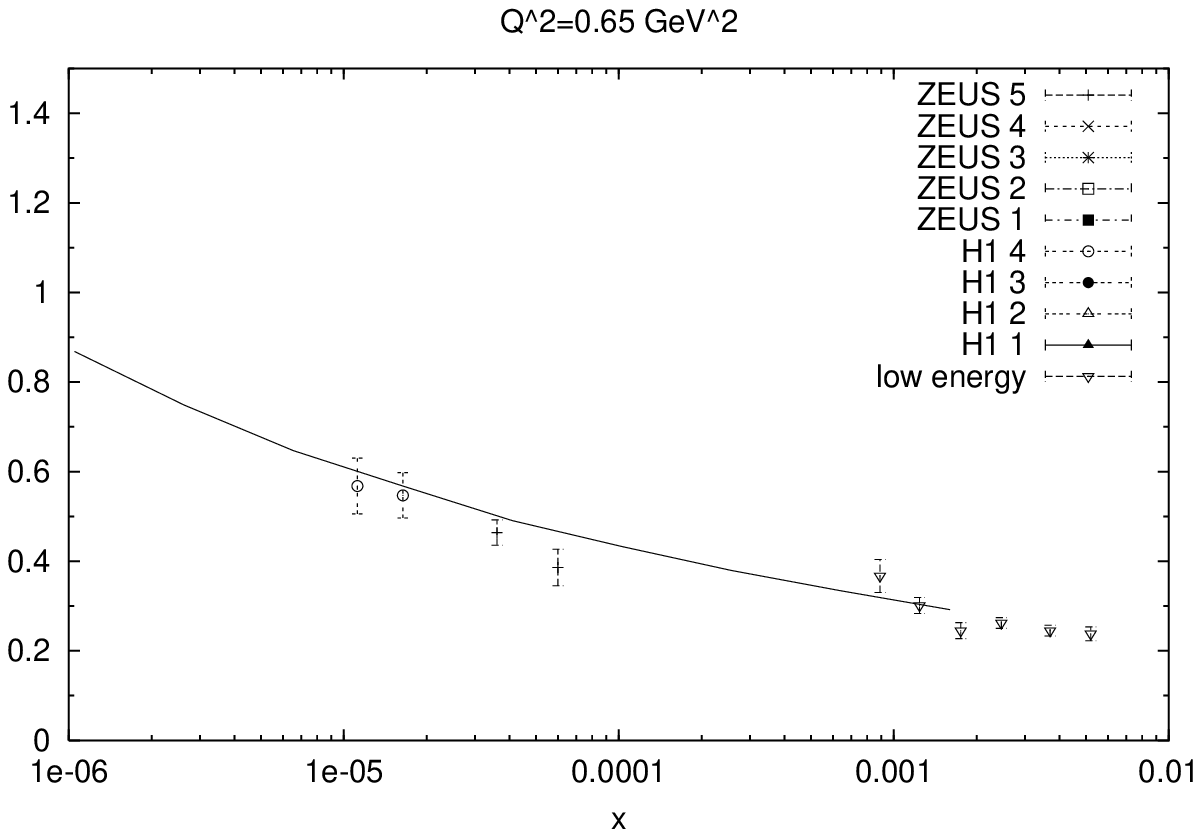}
\end{minipage}
\begin{minipage}{5.2cm}
\includegraphics[width=5.2cm]{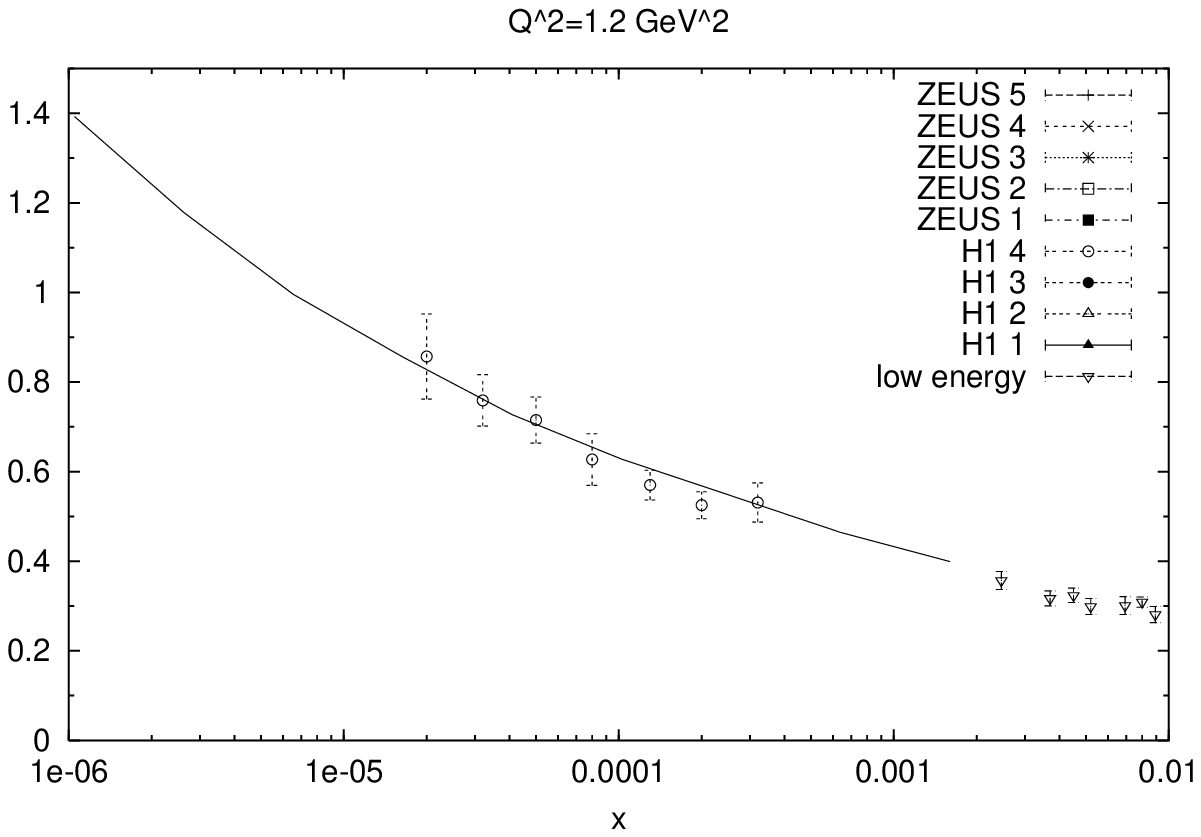}
\end{minipage}
\begin{minipage}{5.2cm}
\includegraphics[width=5.2cm]{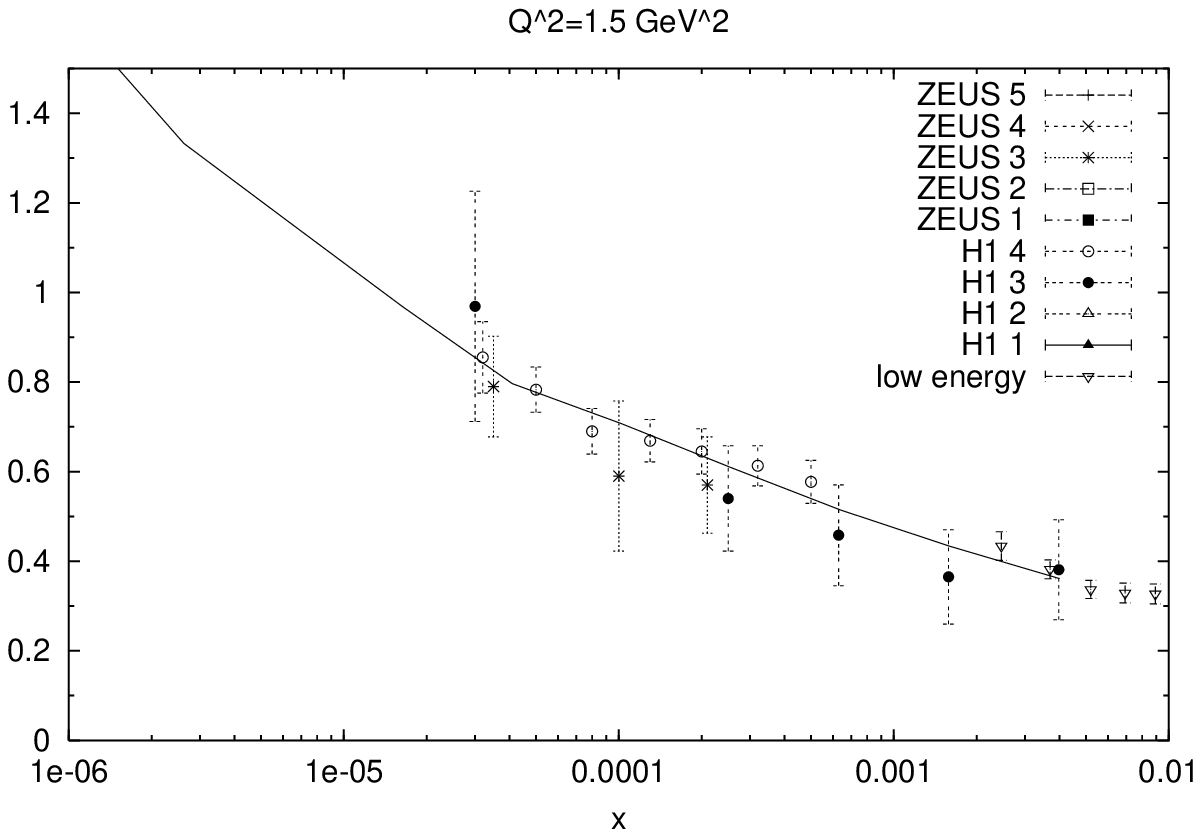}
\end{minipage}
\begin{minipage}{5.2cm}
\includegraphics[width=5.2cm]{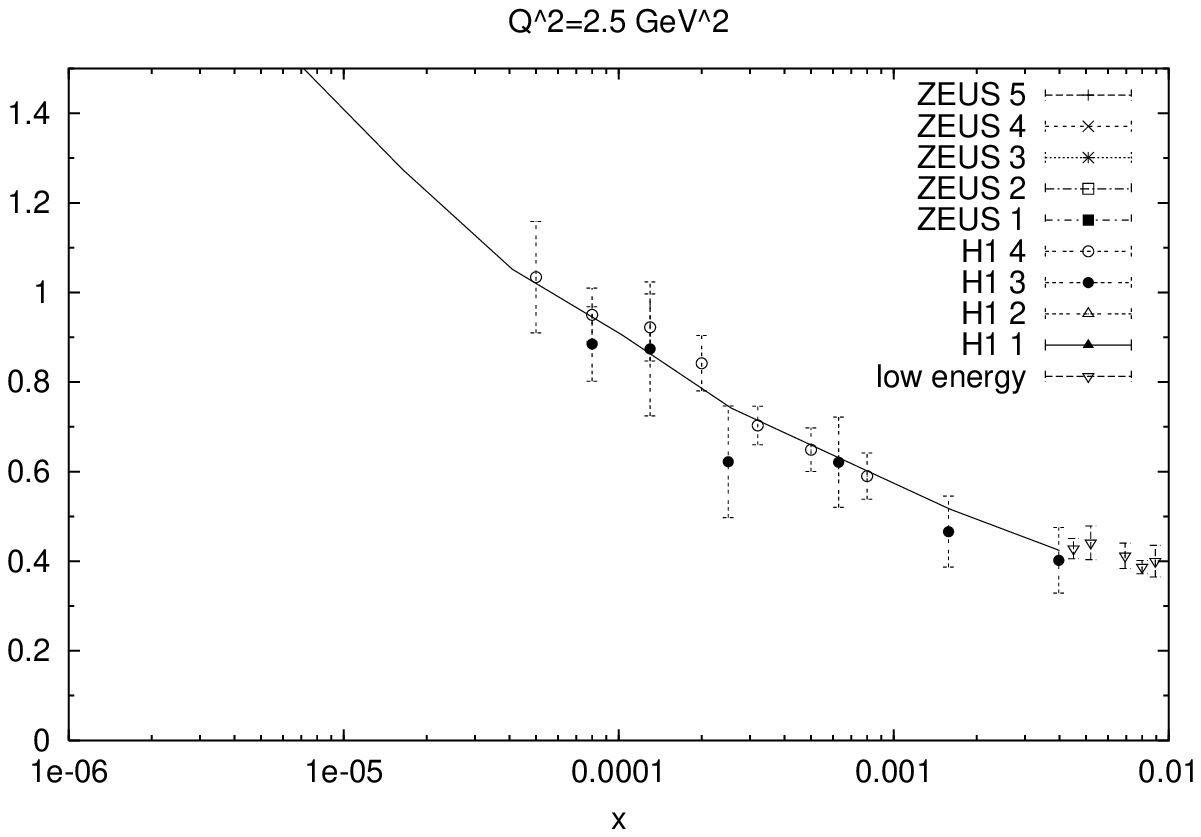}
\end{minipage}
\begin{minipage}{5.2cm}
\includegraphics[width=5.2cm]{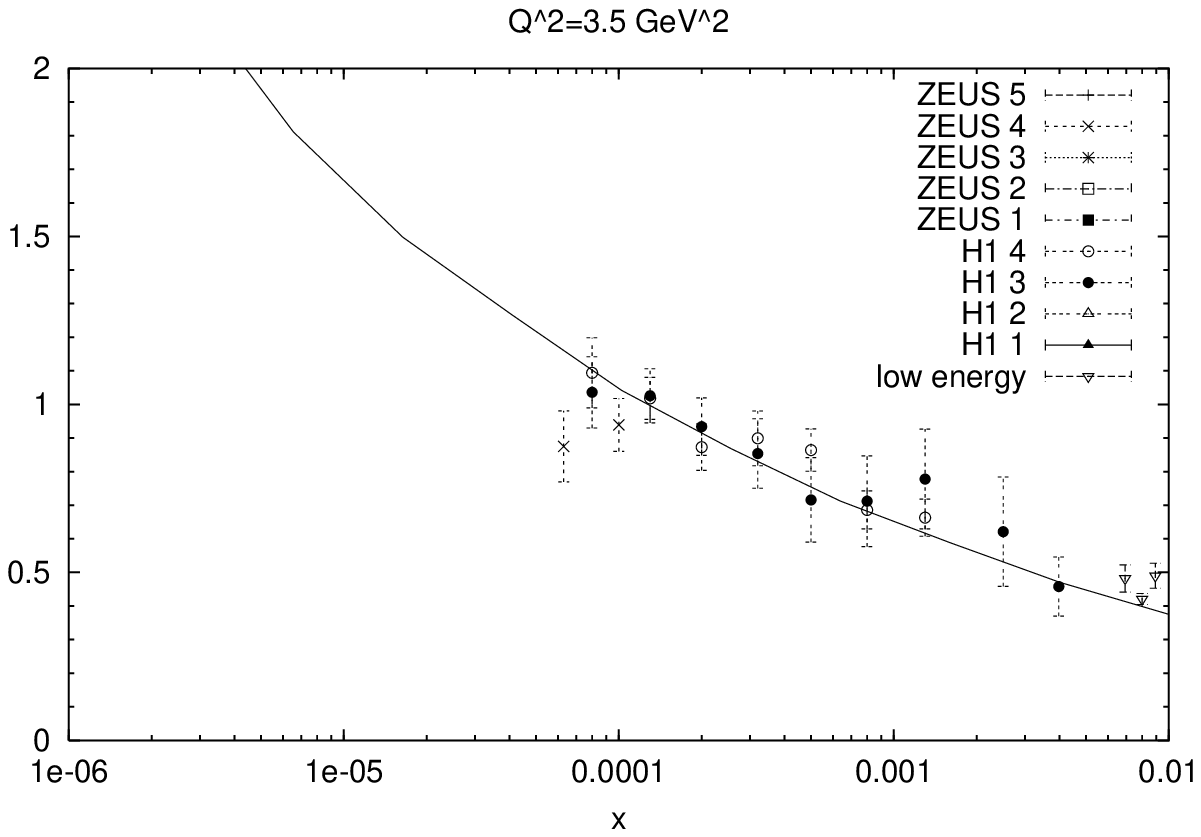}
\end{minipage}
\lbcap{14cm}{The proton structure function $F_2(x,Q^2)$ for fixed values of $Q^2$ as a function of $x$. Here $Q^2\le 3.5 {\rm \, GeV}^2$. The H1 data 1-4 are \protect\cite{Abt:1993,Ahmed:1995,Aid:1996III,Adloff:1997}, the ZEUS data 1-5 are \protect\cite{Derrick:1993,Derrick:1995III,Derrick:1996III,Derrick:1996IIII,Breitweg:1997III} and the low-energy data are \protect\cite{Benvenuti:1989,Arneodo:1997,Adams:1996}.}{F21}
\begin{figure}[ht]
\leavevmode
\begin{minipage}{5.2cm}
\includegraphics[width=5.2cm]{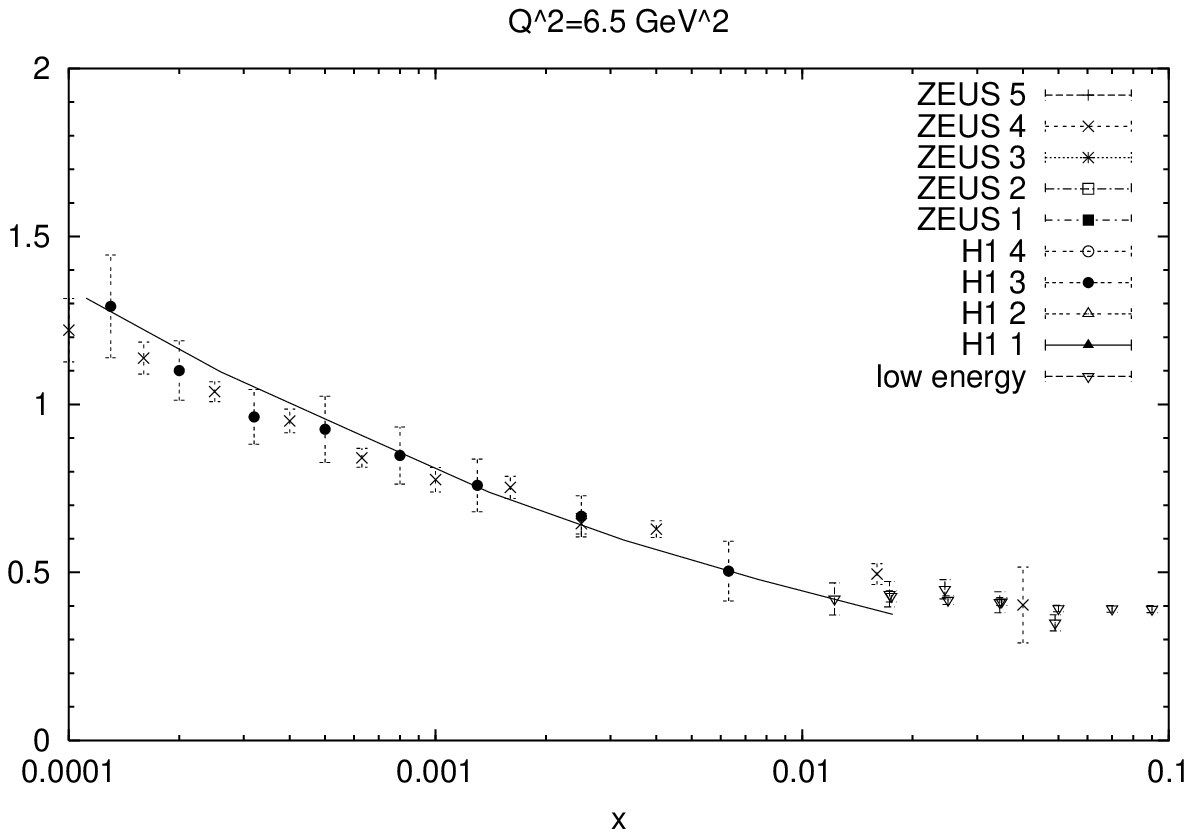}
\end{minipage}
\begin{minipage}{5.2cm}
\includegraphics[width=5.2cm]{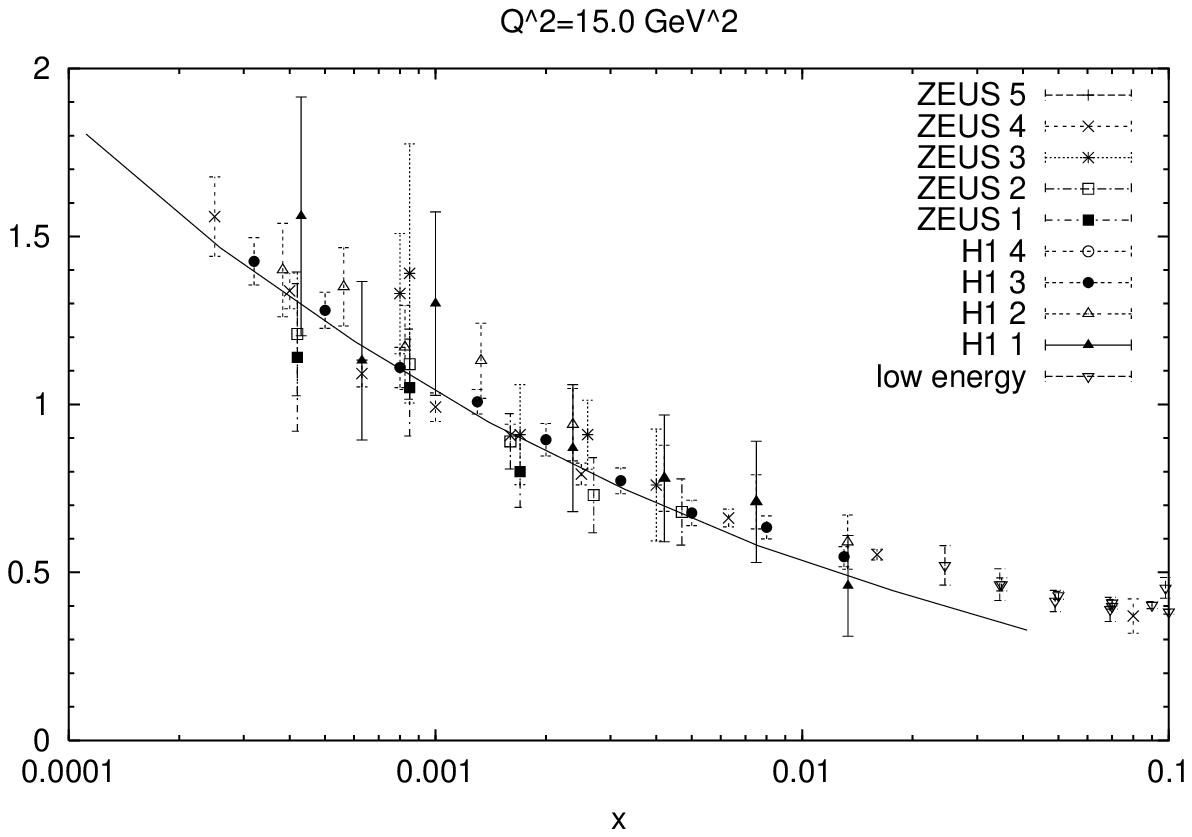}
\end{minipage}
\begin{minipage}{5.2cm}
\includegraphics[width=5.2cm]{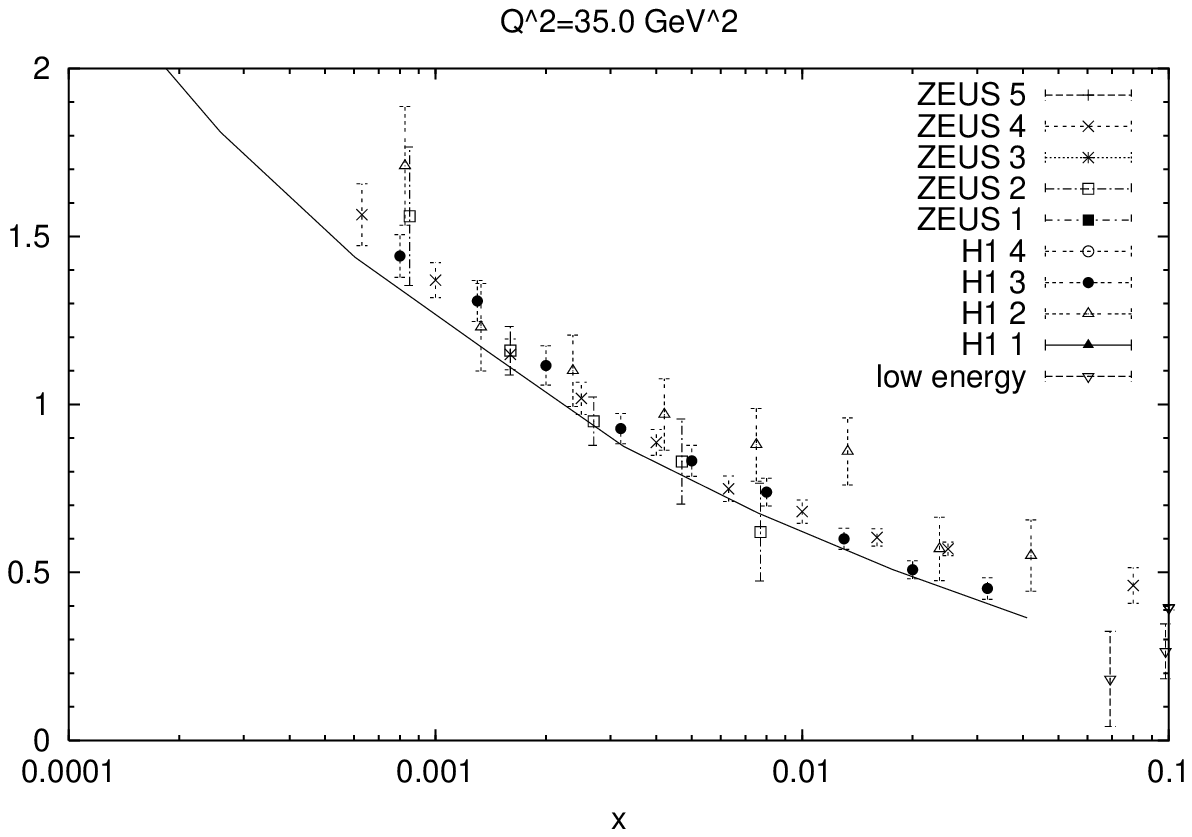}
\end{minipage}
\begin{minipage}{5.2cm}
\includegraphics[width=5.2cm]{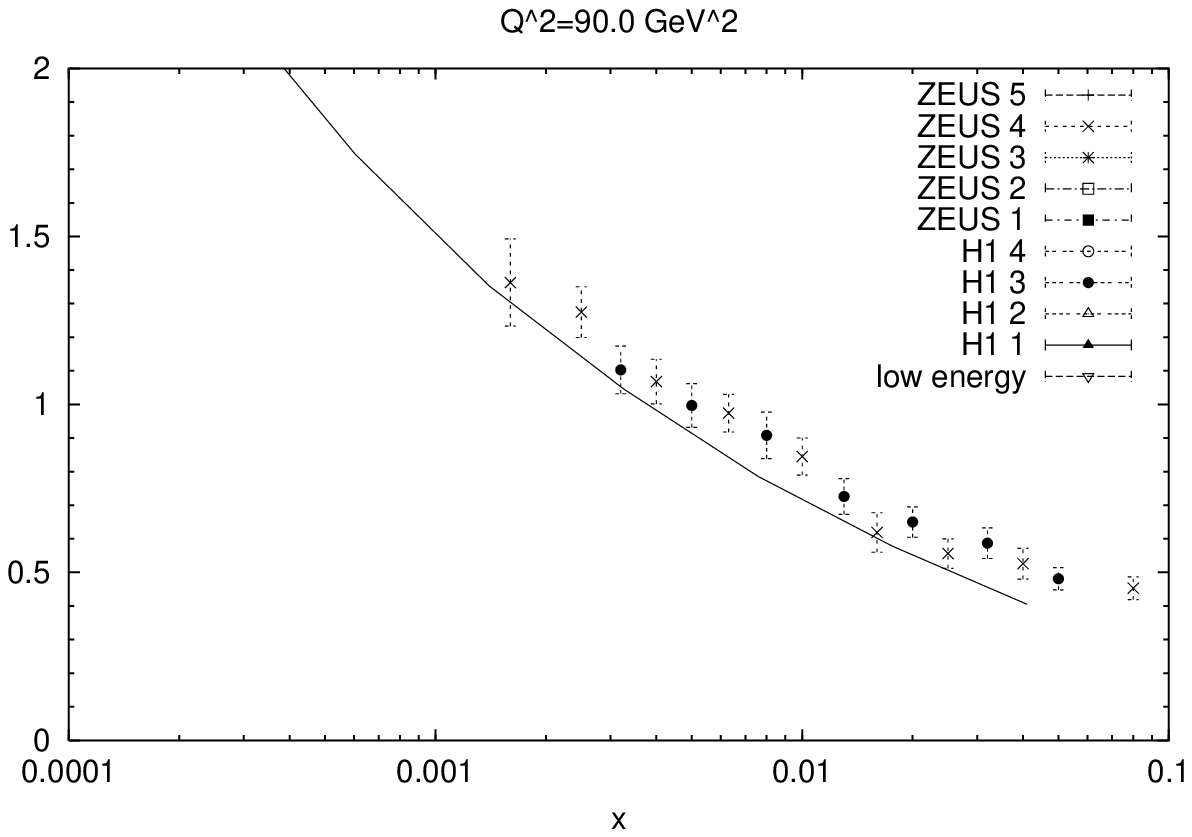}
\end{minipage}
\begin{minipage}{5.2cm}
\includegraphics[width=5.2cm]{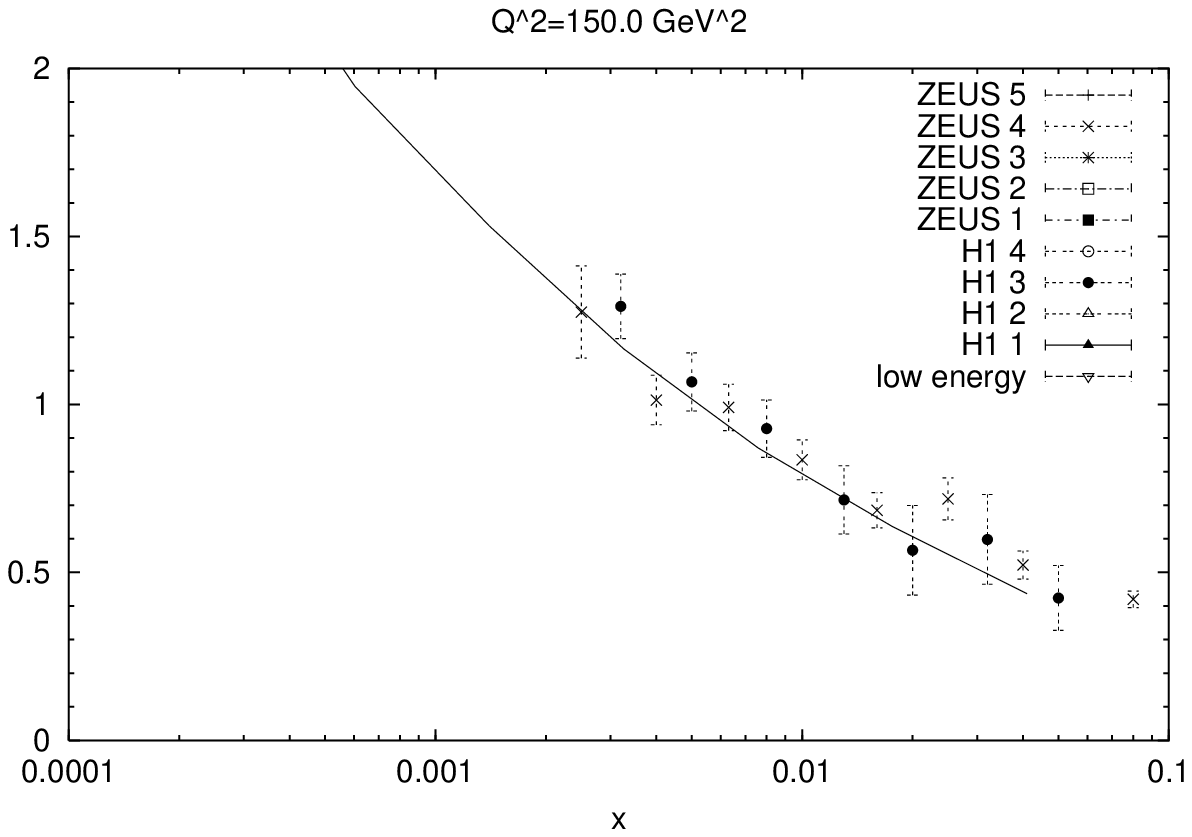}
\end{minipage}
\begin{minipage}{5.2cm}
\includegraphics[width=5.2cm]{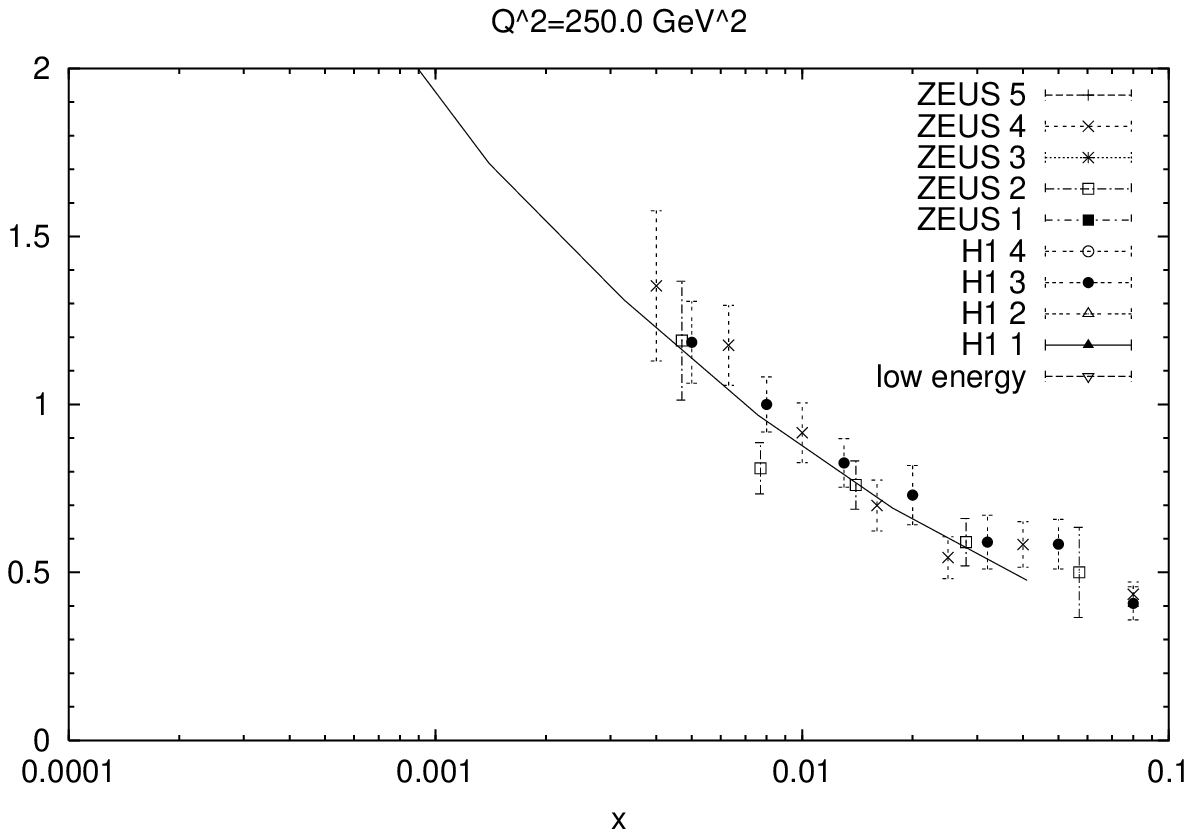}
\end{minipage}
\begin{minipage}{5.2cm}
\includegraphics[width=5.2cm]{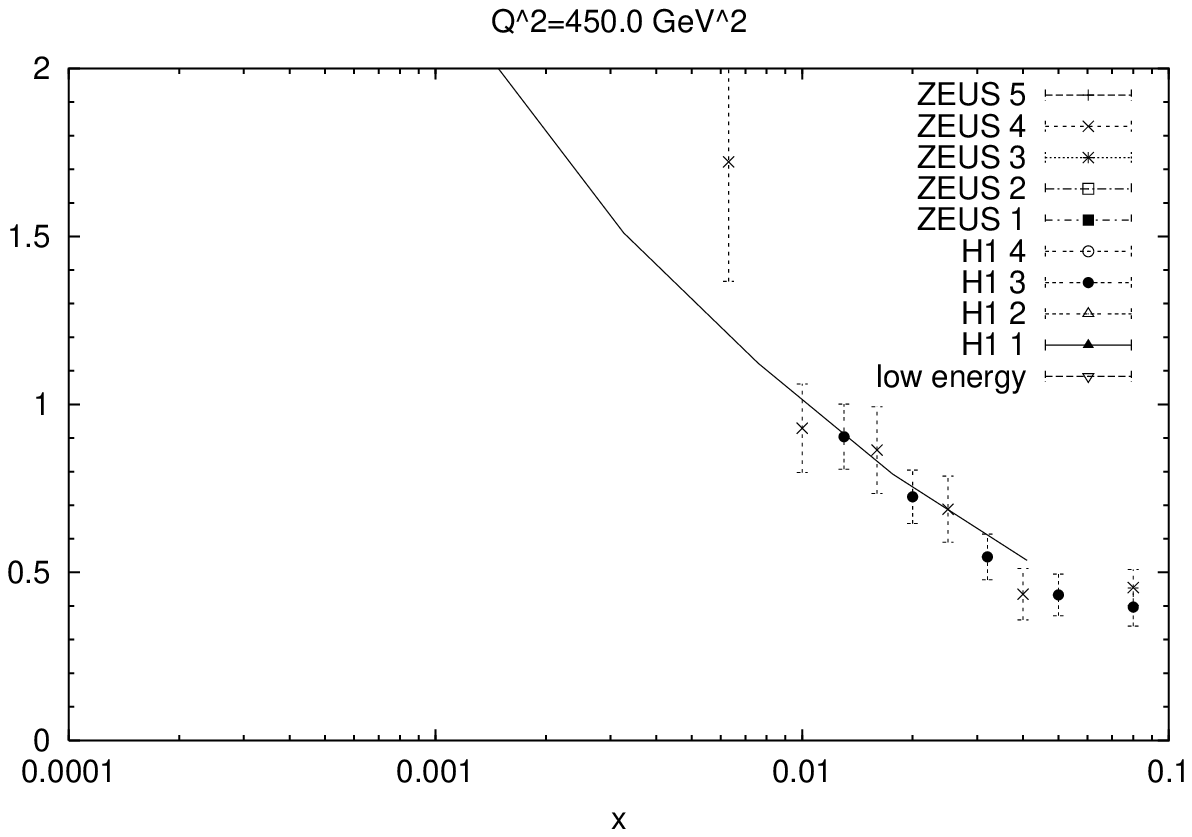}
\end{minipage}
\begin{minipage}{5.2cm}
\includegraphics[width=5.2cm]{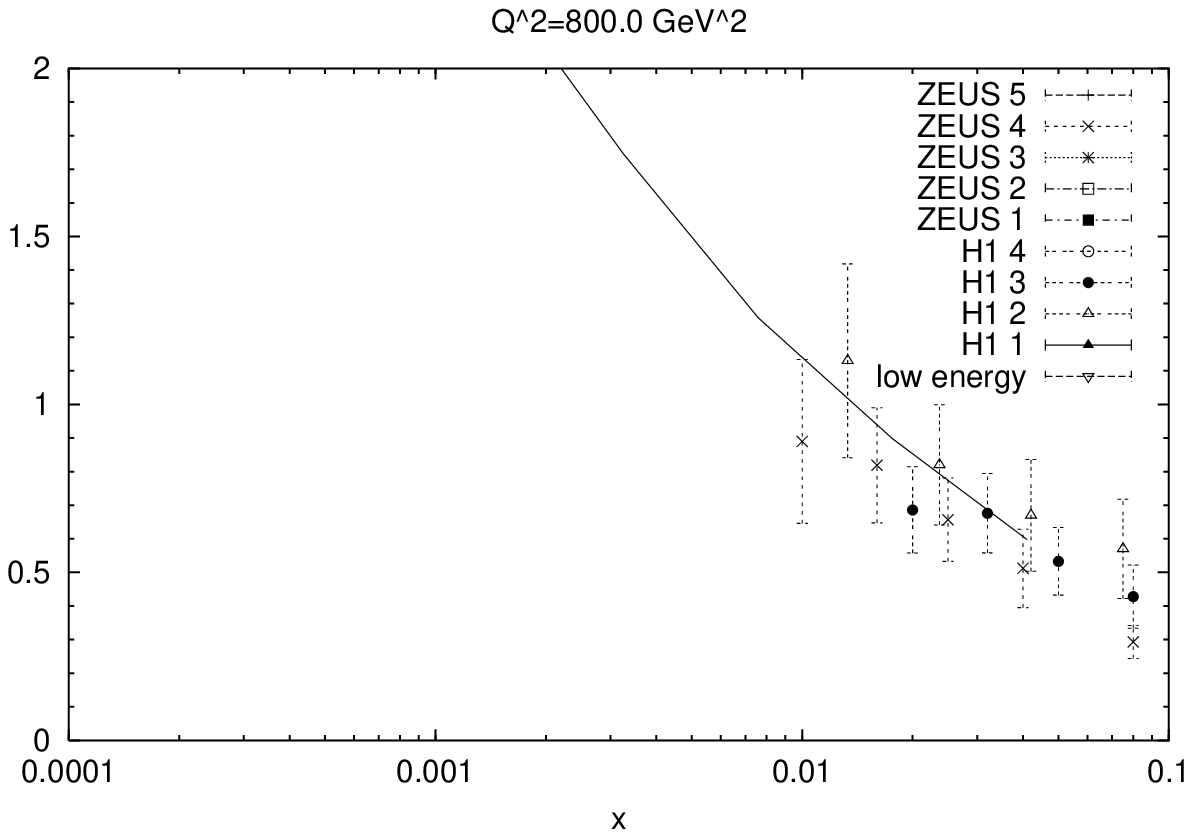}
\end{minipage}
\begin{minipage}{5.2cm}
\includegraphics[width=5.2cm]{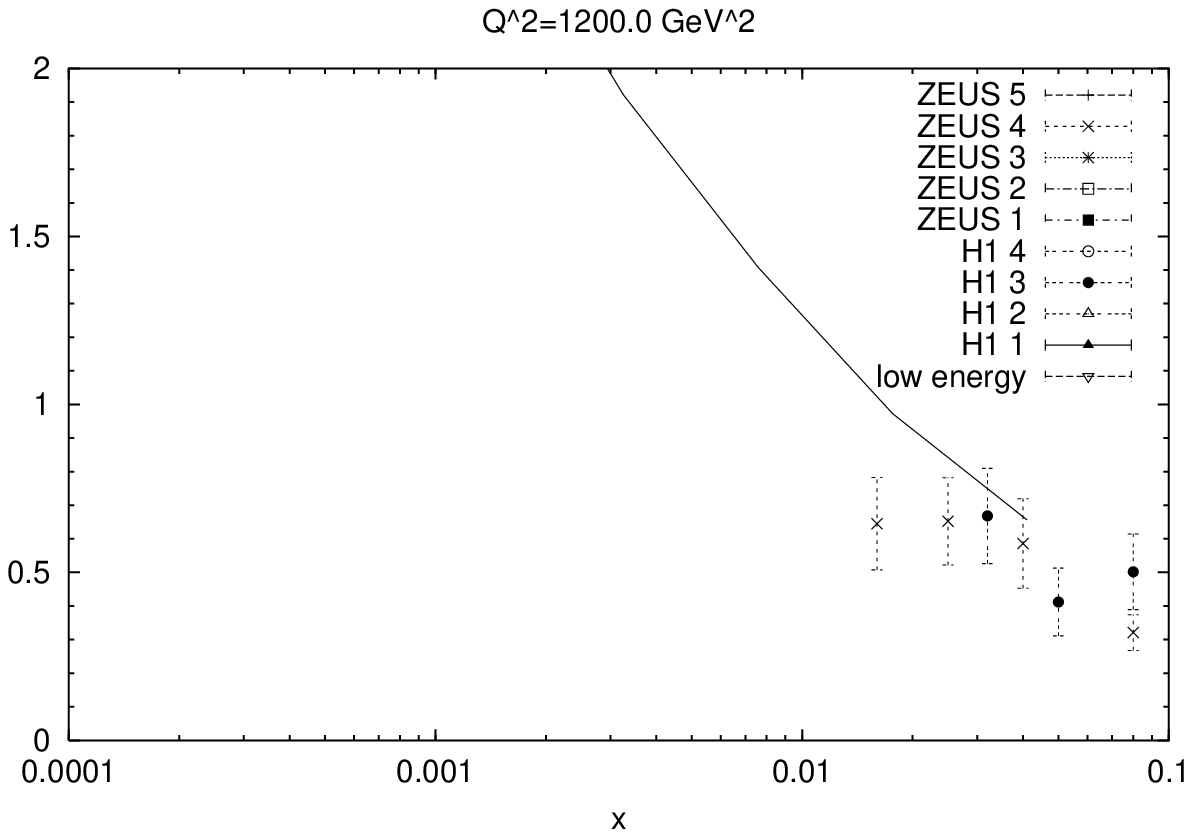}
\end{minipage}
\begin{minipage}{5.2cm}
\includegraphics[width=5.2cm]{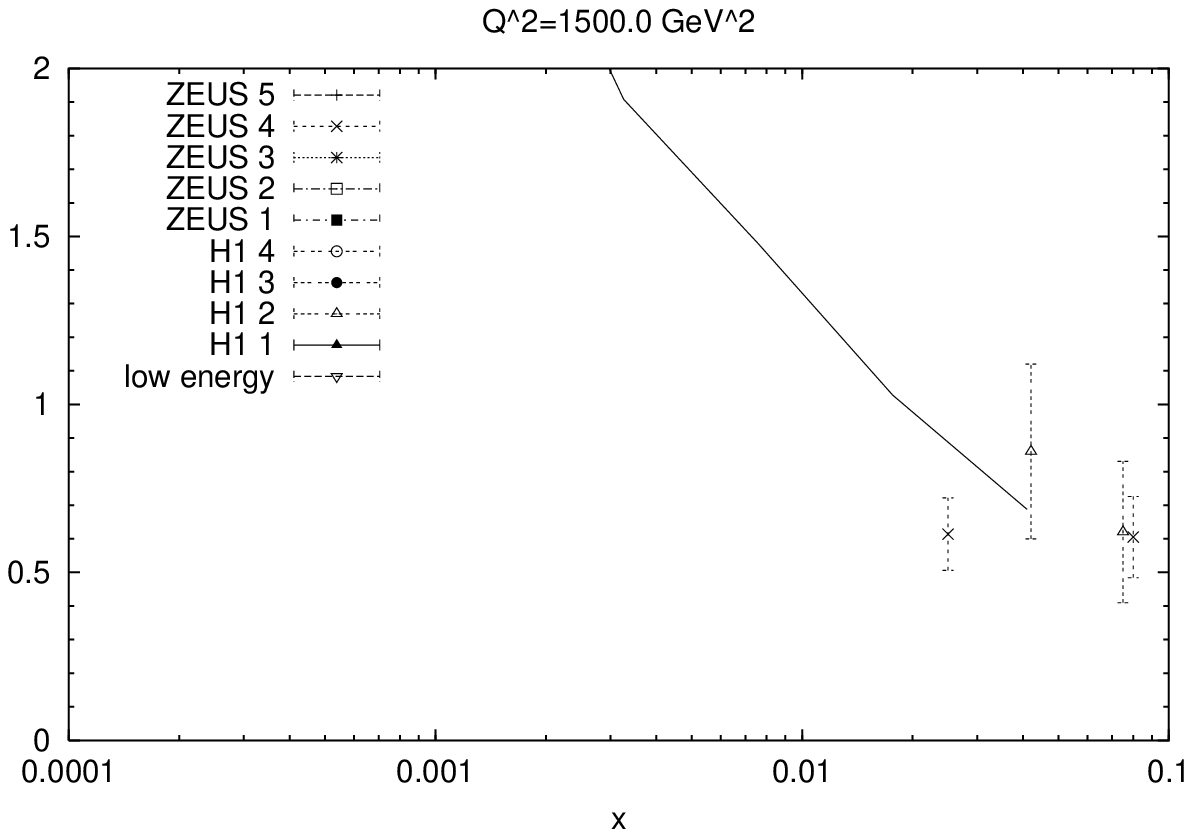}
\end{minipage}
\begin{minipage}{5.2cm}
\includegraphics[width=5.2cm]{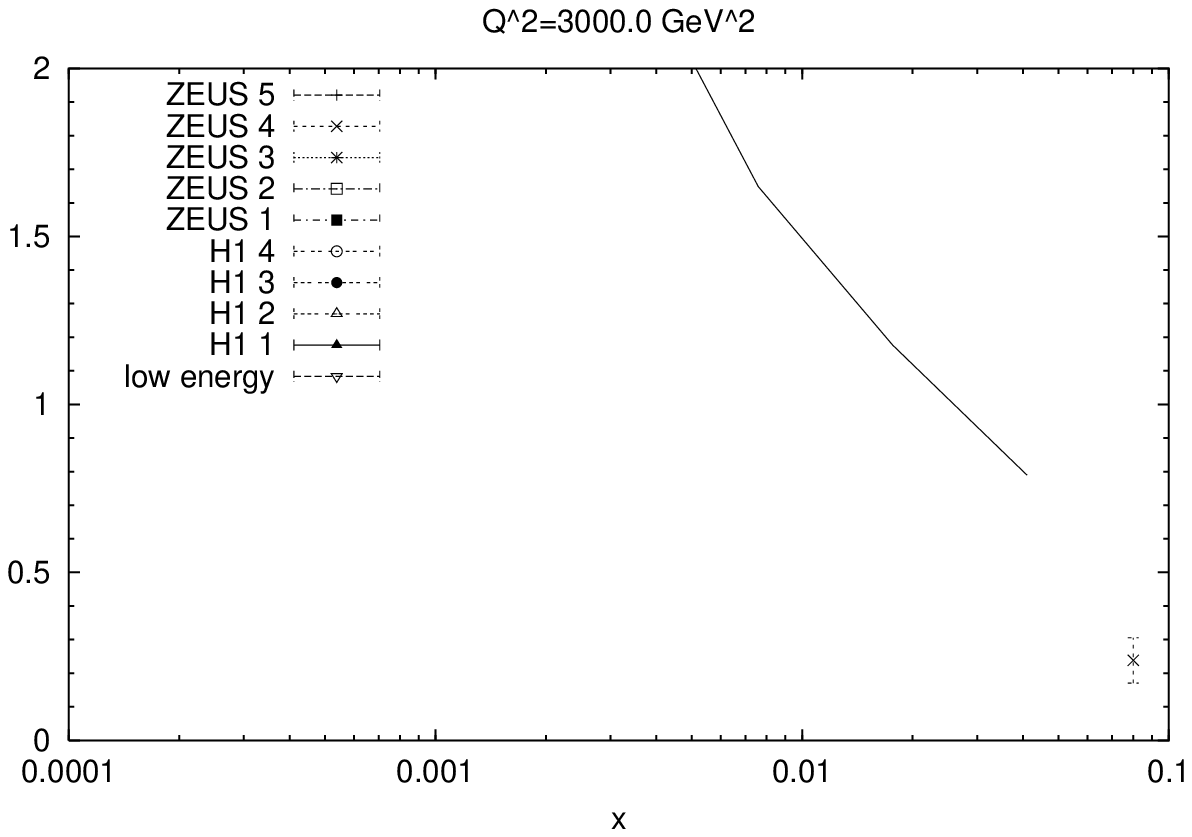}
\end{minipage}
\begin{minipage}{5.2cm}
\includegraphics[width=5.2cm]{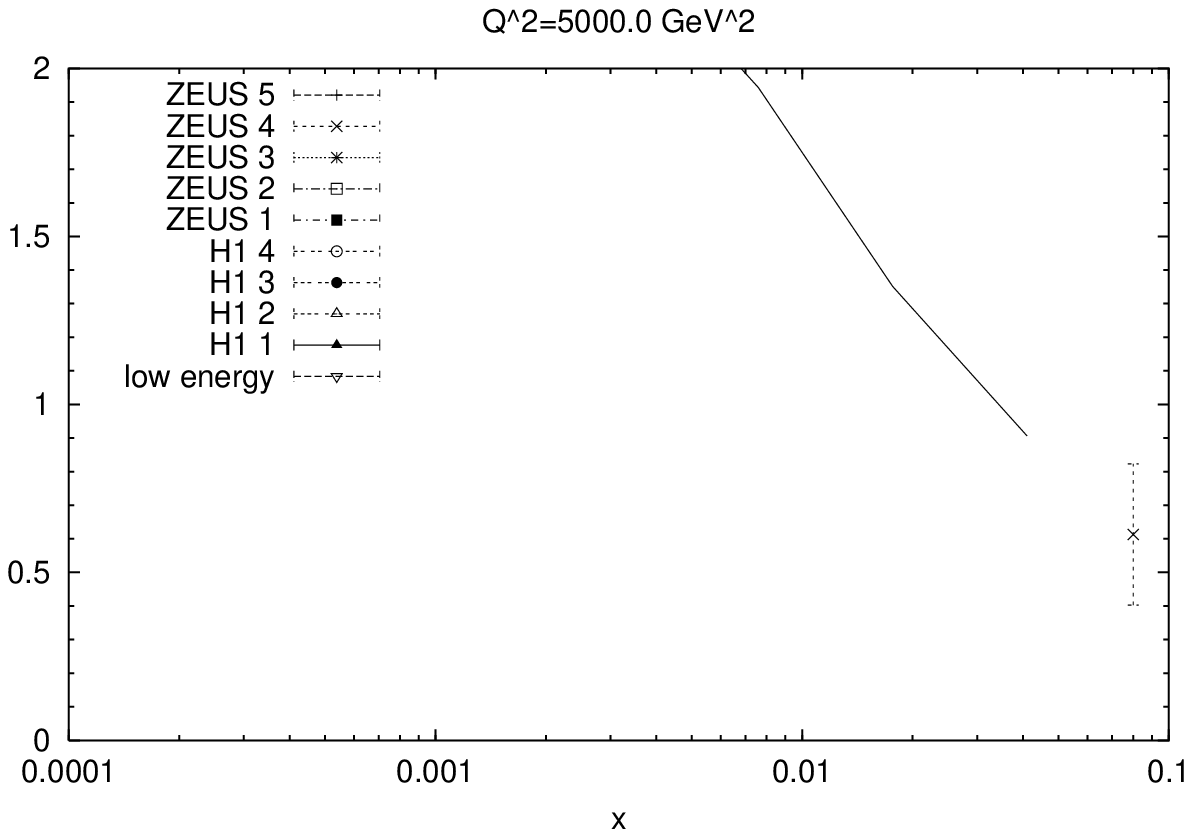}
\end{minipage}
\lbcap{14cm}{The proton structure function $F_2(x,Q^2)$ for fixed values of $Q^2$ as a function of $x$. Here $Q^2> 3.5 {\rm \, GeV}^2$. The data set is the same as in \fig{F21}.}{F22}
\befi{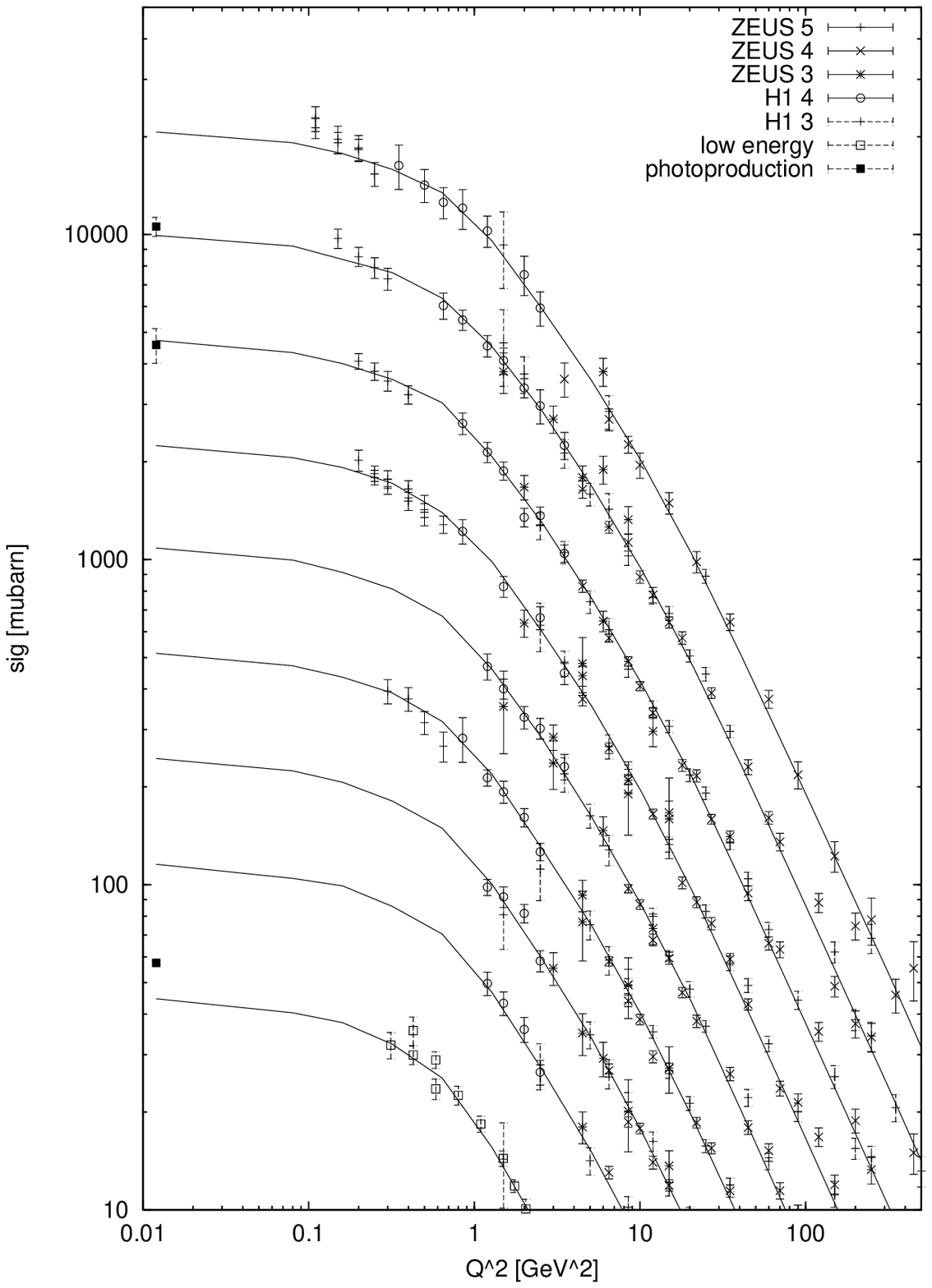}{13cm}
\unitlength1cm
\begin{picture}(0,0)
\footnotesize
\put(3.7,3.8){$W=20\GeV\;(\times\;0.5)$}
\put(3.7,6.0){$W=60\GeV\;(\times\;1)$}
\put(3.7,7.5){$W=75\GeV\;(\times\;2)$}
\put(3.7,8.8){$W=95\GeV\;(\times\;4)$}
\put(3.7,10.3){$W=115\GeV\;(\times\;8)$}
\put(3.7,11.6){$W=140\GeV\;(\times\;16)$}
\put(3.7,13.0){$W=170\GeV\;(\times\;32)$}
\put(3.7,14.5){$W=210\GeV\;(\times\;64)$}
\put(3.7,15.9){$W=245\GeV\;(\times\;128)$}
\end{picture}
\lbcap{14cm}{The total cross section. The data set is the same as in \fig{F21}. For the data selection we allowed $W$ values which differ by at most 10\% from the quoted value. The photoproduction data are from \protect\cite{Caldwell:1978,Aid:1995,Derrick:1994} and are displayed at $Q^2=0.01{\rm\;GeV}^2$ together with our result. For convenience the data are scaled with the given factors.}{sigtotWfest}
\begin{figure}[ht]
\leavevmode
\begin{minipage}{7.3cm}
\includegraphics[width=7.3cm]{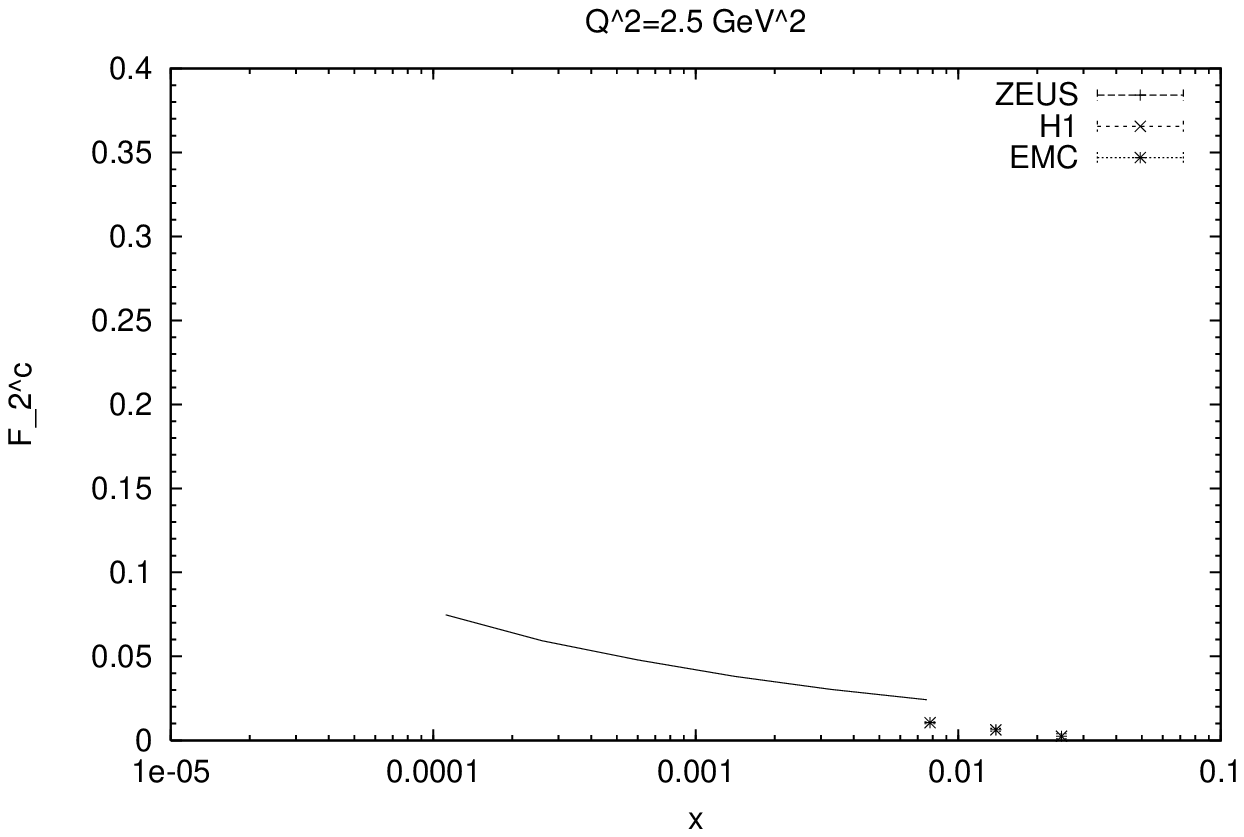}
\end{minipage}
\hfill
\begin{minipage}{7.3cm}
\includegraphics[width=7.3cm]{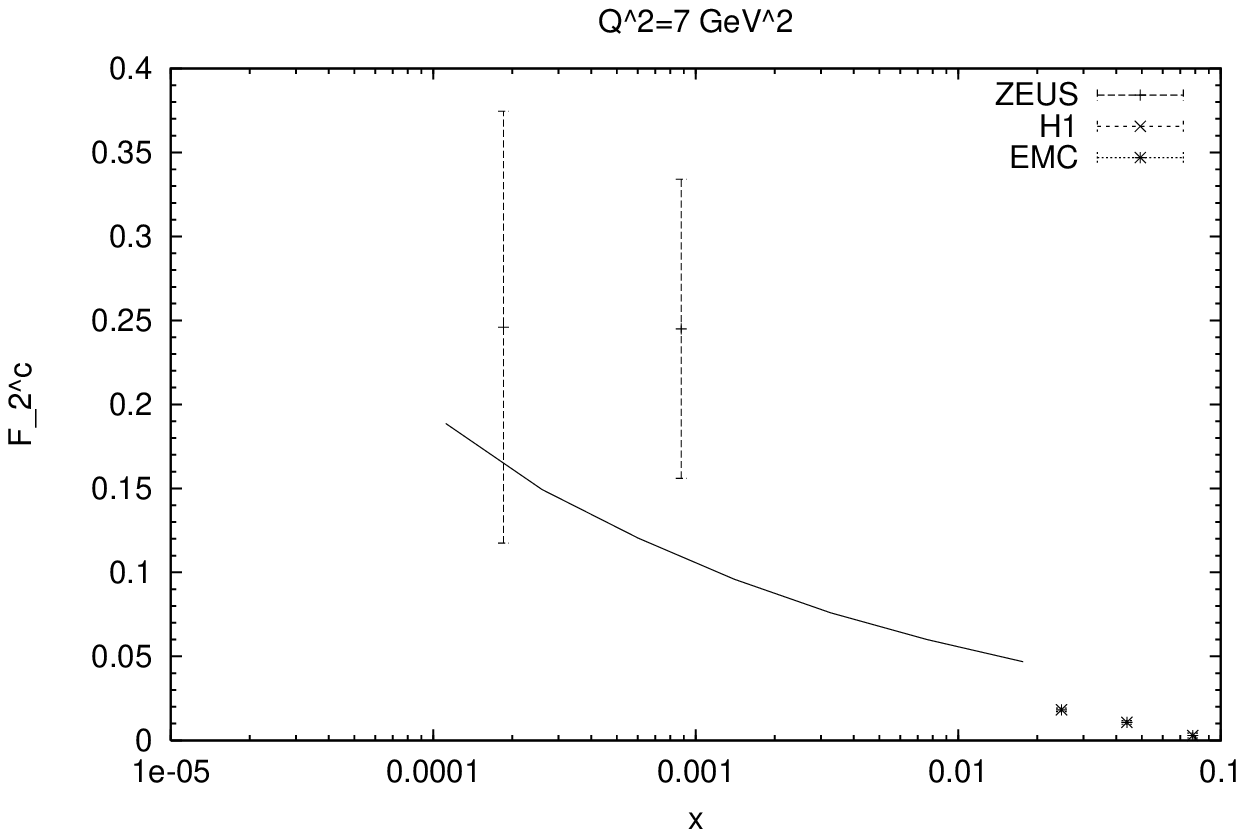}
\end{minipage}
\begin{minipage}{7.3cm}
\includegraphics[width=7.3cm]{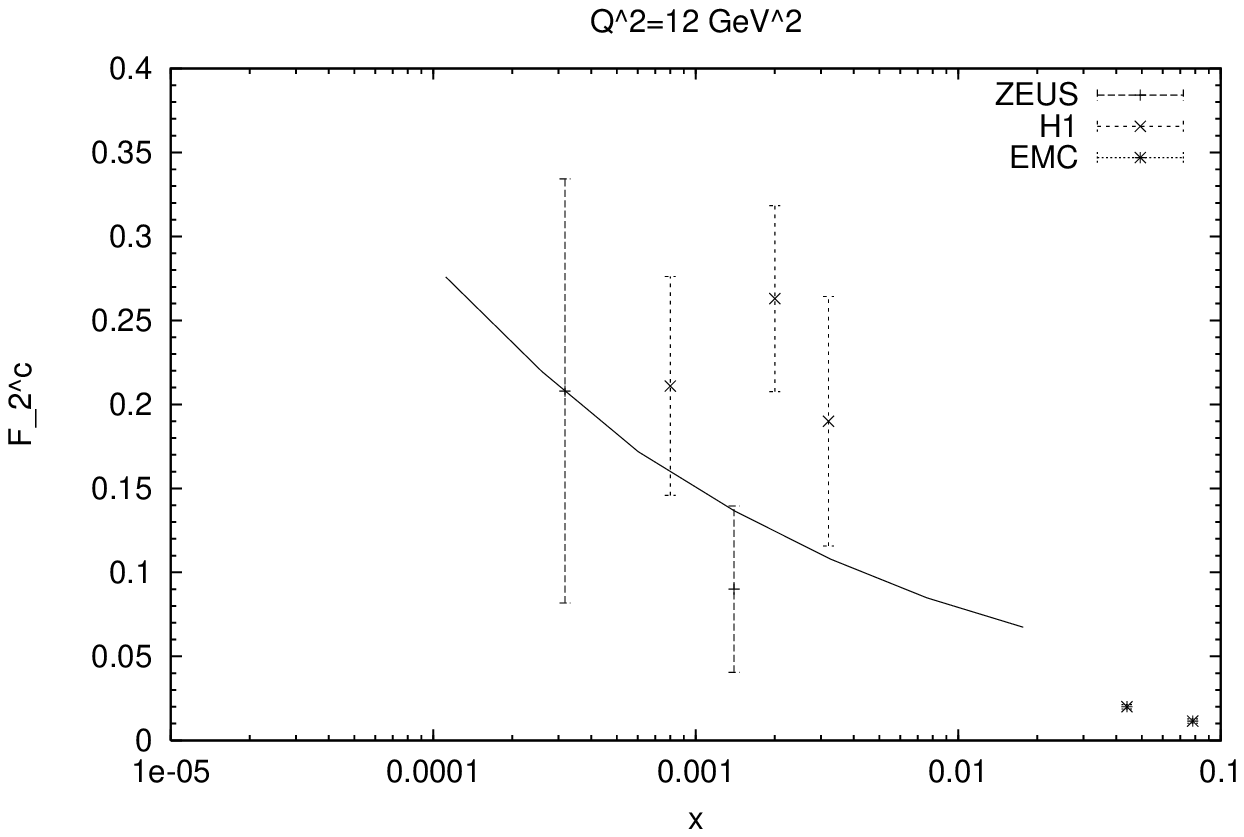}
\end{minipage}
\hfill
\begin{minipage}{7.3cm}
\includegraphics[width=7.3cm]{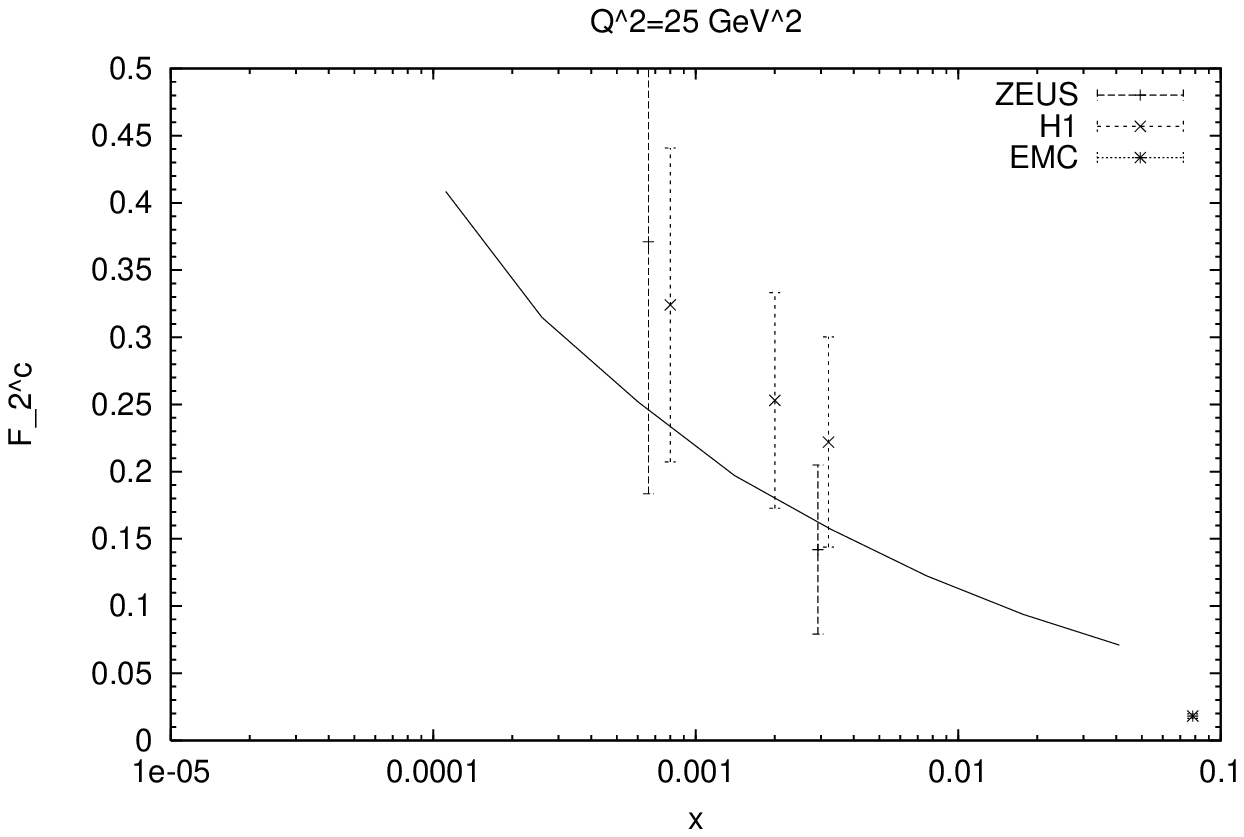}
\end{minipage}
\begin{minipage}{7.3cm}
\includegraphics[width=7.3cm]{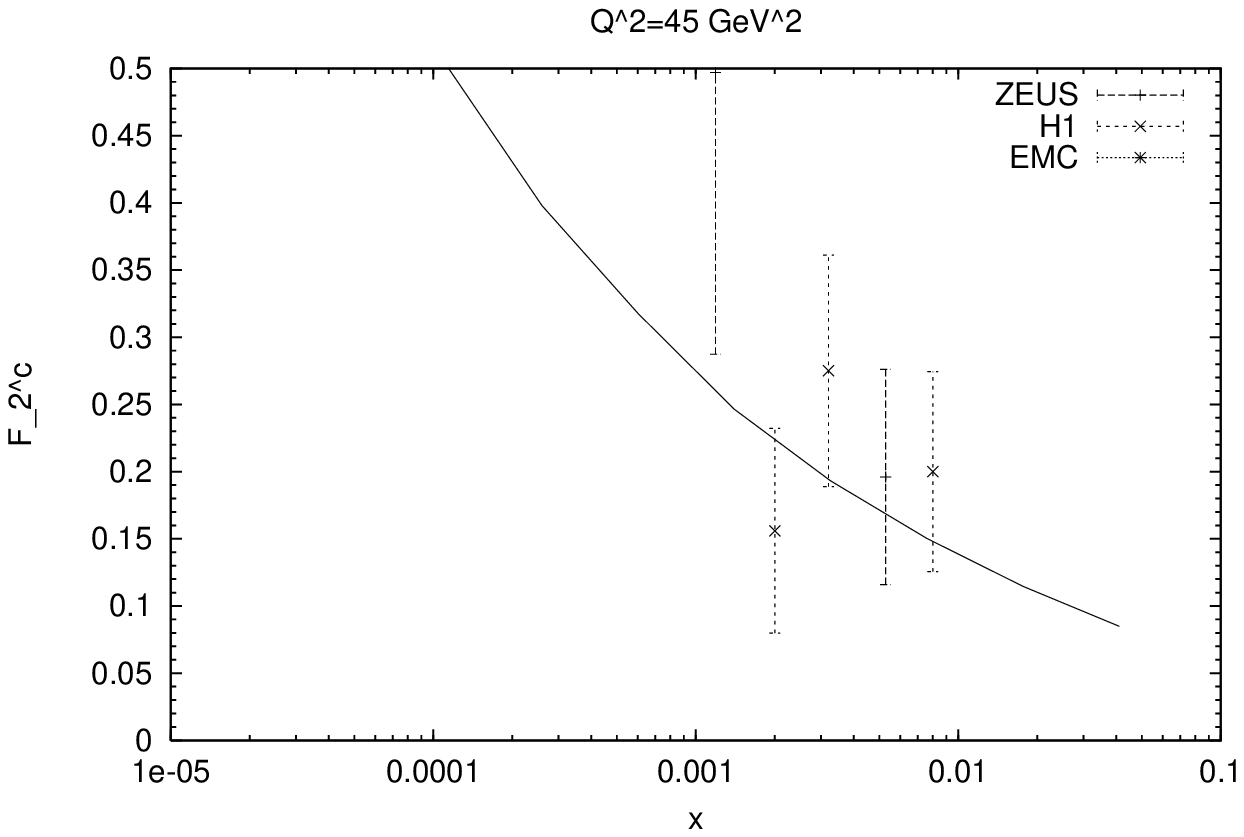}
\end{minipage}
\hfill
\begin{minipage}{7.3cm}
\includegraphics[width=7.3cm]{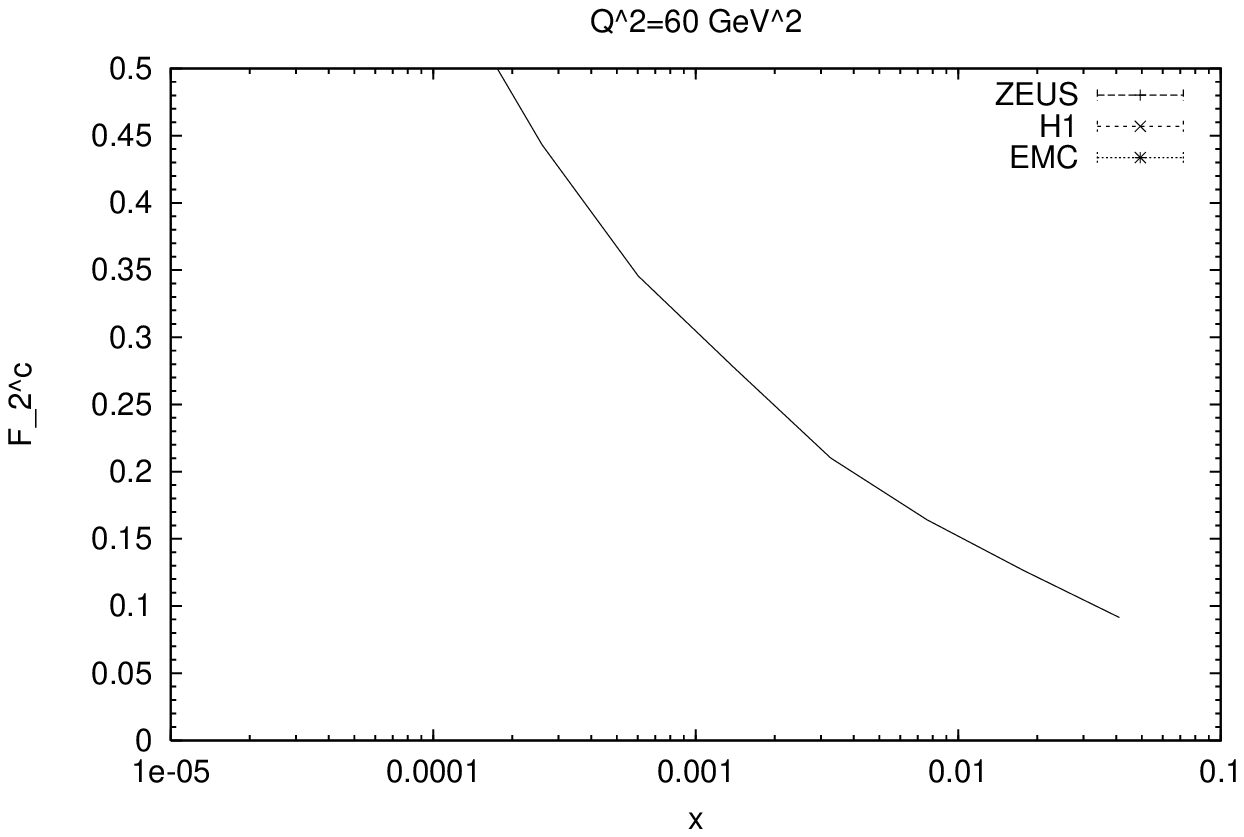}
\end{minipage}
\begin{minipage}{7.3cm}
\includegraphics[width=7.3cm]{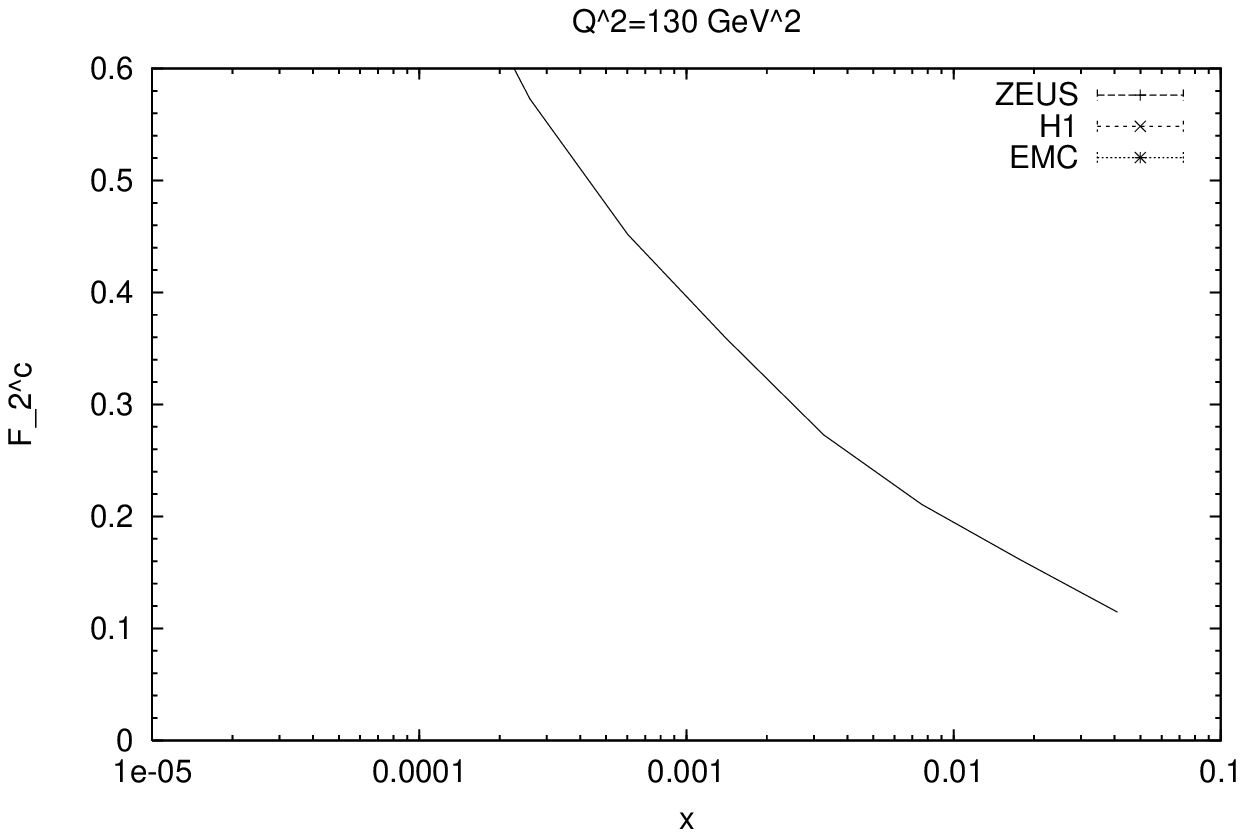}
\end{minipage}
\lbcap{14cm}{The charm contribution $F_2^c(x,Q^2)$ to the proton structure function for fixed values of $Q^2$ as a function of $x$. The H1 data are from reference \protect\cite{Adloff:1996}, the ZEUS data from \protect\cite{Breitweg:1997IIII} and the low-energy data from \protect\cite{Aubert:1983}. There are preliminary ZEUS results (see for example \protect\cite{Bailey:1998}) which also fit our data.}{F2c}
\clearpage
\section{Summary}
In this paper we have demonstrated that the different energy and $Q^2$
behavior of the different considered processes can be described by
making a uniform phenomenological ansatz for the energy dependence of
dipole-dipole scattering from which all processes are constructed. The
energy dependence is due to the exchange of two pomerons between the
dipoles. The soft-pomeron with an intercept of 1.08 couples only to
dipoles which are larger than the cut $c$ and the hard-pomeron with an
intercept of 1.28 contributes if at least one dipole is smaller than
$c$. For the slope of the soft-pomeron we take the standard value of
$0.25\GeV^{-2}$ whereas we assume the slope of the hard-pomeron to be
zero at least for small $t$. The main goal of this paper is not to
present a fit of $F_2$ or the vectormeson production data but to show
how the different effective energy behavior of the different processes
is due to the wavefunctions making the process dominated by smaller or
larger dipoles.

Our approach, which turns out to describe all these processes well is
based on the following assumptions: The processes can be calculated by
smearing the dipole-dipole scattering with appropriate wavefunctions
which are either phenomenologically (hadrons, vectormesons) or
perturbatively motivated (photons). At fixed cm-energy of $20 \GeV$
the dipole-dipole scattering can be calculated using the Model of the
Stochastic Vacuum. This nonperturbative model can only be used if the
dipoles are not to small. To describe the data up to $Q^2\le 35\GeV^2$
it is sufficient just to cut dipoles which are smaller than a new
introduced cut $r_{\rm cut}$. In this kinematic regime one observes
the transition from the soft to the hard behavior and this can be
described in our model very well. We then showed that we can extend
our approach to even harder processes by calculating for the very
small dipoles the leading perturbative contribution with a running
strong coupling on the 1-loop level which is frozen in the infra-red
to be $\al_s(\infty)$. But for such hard processes the more
sophisticated perturbative descriptions work very well and this is not
the regime of our main interest.

By adjusting the three new parameters ($c$, $r_{\rm cut}$, $\al_s(\infty)$) we obtain a very good description of the experimental results for the following physical values
\[
\begin{tabular}{|c|c|c|}
\hline
$c$&$r_{\rm cut}$&$\al_s(\infty)$\\
\hline\hline
0.35 fm&0.16 fm&0.75\\
\hline
\end{tabular}
\]
We want to point out that the obtained values are not the main result
of this paper. Indeed by changing for example slightly the
hard-pomeron intercept one obtains after readjusting the three
parameters a quite similar good fit.  In this framework we obtain not
only the right transition from the soft to the hard energy dependence
but do also predict the absolute size of the cross sections.
Especially we want to point out that we get simultaneously the strong
energy dependence of photoproduction of the $J/\Ps$ and the small-$x$
dependence of $F_2$ for all values of $Q^2$.

Off course we have also limitations in our approach: The considered
cm-energy has to be large enough to ensure that no Regge-trajectories
are important. We can only consider soft processes where the internal
energy is the largest scale. Thus our $x$ values are limited to be
small enough or for electroproduction with large $Q^2$ we have to go
to larger $W$.

One important observation in our approach is that we have to couple
quite large dipoles (up to $0.35\fm$) to the hard-pomeron. The
scattering of these dipoles can not be calculated in a simple
perturbative way and we use the Model of the Stochastic Vacuum.
Exactly these dipoles are responsible for the strong rise of the
$J/\Ps$ photoproduction with $W$.

Our treatment of the energy dependence due to the two pomerons is very
similar to the recent publication of Donnachie and Landshoff
\cite{Donnachie:1998}. Whereas DL had to fit the coupling as a
function of $Q^2$ to the HERA data we couple the pomerons to the
dipoles and their interaction is calculated as described above. The
main difference of these two approaches is that in our approach the
structure function $F_2$ can not be written as
$F_2=a\,x^{-0.08}+b\,x^{-\ep_{\rm hard}}$ because off the additional
$W$ dependence of the photon wavefunction. DL obtain a very large
intercept for the hard-pomeron, $\ep_{\rm hard}=0.435$, whereas our
$\ep_{\rm hard}$ is 0.28. In their paper DL point out that also their
hard-pomeron intercept is not very well fixed by the fit and has a
large error. Both approaches can describe the data because in our
treatment the effective power is enhanced due to the photon
wavefunction whereas DL obtain a very large contribution from the
soft-pomeron even for $Q^2\ge 100\GeV^2$ which makes the effective
power smaller.

Maybe the nicest feature of the presented approach is, that one can
calculate the energy dependence of all processes based on
dipole-dipole scattering without any new parameters. In reference
\cite{Donnachie:1998II} we investigate for example the $\ga$-$\ga$
physics and the results are very satisfactory without any free
parameters.
\section*{Acknowledgments}
I would like to thank H.G.~Dosch and Sandi Donnachie for many fruitful
discussions, suggestions and for reading the manuscript. I am
especially grateful to the {\it Minerva}-Stiftung for my fellowship.
This work started during a stay in Heidelberg and I want to thank the
theory-group for their hospitality and the {\it Graduiertenkolleg} for
financial support.
\section*{Appendix: The photon and vectormeson wavefunctions}
For longitudinal polarized photon we have
\beq
\Psi^{\ga}_{f h_1 h_2}(\vec{r}_\ga,z)=-\sqrt{N_{\rm C}} e_f \de_{h_1, -h_2}2z(1-z)Q\frac{K_0(\ep r_\ga)}{2\pi}
\lbq{gammalongwf}
where $\ep=\sqrt{z(1-z)Q^2+m_f(Q^2)}$ and $e_f$ is the quark charge. In reference \cite{Dosch:1997II} the application was extended to real photons by using (anti)quark masses that depend on the virtuality and become equal to the constituent masses for $Q^2=0$. We use in this paper the parameterization given in reference \cite{Dosch:1997II}, eq.~18/19:
\beqa
m_{u,d}&=&\left\{ {0 \; | \; Q^2> 1.05 \GeV^2} \atop{0.22 \GeV(1-Q^2/1.05\GeV^2)\;| \;Q^2 \le 1.05 \GeV^2}\right\}\nn\\
m_s&=&\left\{ {0.15\GeV \; | \; Q^2> 1.6 \GeV^2} \atop{0.15\GeV + 0.16 \GeV(1-Q^2/1.6\GeV^2)\;| \;Q^2 \le 1.6 \GeV^2}\right\}\nn\\
m_c&=&1.3\GeV.
\lbqa{runningmf}
For transversal photons, e.g.~with polarization $\la = +$, we obtain:
\beq
\Psi^{\ga}_{f h_1 h_2}(\vec{r}_\ga,z)=\sqrt{N_{\rm C}} e_f \sqrt 2 \left( i e^{i\th}\ep\left(z\de_{+-}-(1-z)\de_{-+}\right)\frac{K_1(\ep r_\ga)}{2\pi}+m_f(Q^2)\de_{++}\frac{K_0(\ep r_\ga)}{2\pi}\right)
\lbq{gammatranswf}
where $\th$ is the angle of $\vec r_\ga$ in polar coordinates and $\de_{+-}=\de_{h_1,+}\de_{h_2, -}$. For a transversal photon with $\la=-$ we find analogously
\beq
\Psi^{\ga}_{f h_1 h_2}(\vec{r}_\ga,z)=\sqrt{N_{\rm C}} e_f \sqrt 2 \left( i e^{-i\th}\ep\left((1-z)\de_{+-}-z\de_{-+}\right)\frac{K_1(\ep r_\ga)}{2\pi}+m_f(Q^2)\de_{--}\frac{K_0(\ep r_\ga)}{2\pi}\right).
\lbq{gammatranswf2}
For transversal vectormesons with $\la = +$ we obtain
\beqa
&&\Psi^{\rm VM}_{f h_1 h_2}(\vec{r},z)=\\
&&\frac{c_f^{\rm VM}}{\sum_{f'} c_{f'}^{\rm VM}e_{f'}/e}\left( \frac{i\om^2re^{i\th}}{M_{\rm VM}}\left( z \de_{+-}-(1-z)\de_{-+}\right)+\frac{m_f(Q^2)}{M_{\rm VM}} \de_{++}\right)\frac{\sqrt 2 \pi f^{\rm VM}}{\sqrt{N_{\rm C}}}f(z)e^{-\om^2 r^2/2}\nn
\lbqa{VMtranswf}
and for $\la=-$
\beqa
&&\Psi^{\rm VM}_{f h_1 h_2}(\vec{r},z)=\\
&&\frac{c_f^{\rm VM}}{\sum_{f'} c_{f'}^{\rm VM}e_{f'}/e}\left( \frac{i\om^2re^{-i\th}}{M_{\rm VM}}\left( (1-z) \de_{+-}-z\de_{-+}\right)+\frac{m_f(Q^2)}{M_{\rm VM}} \de_{--}\right)\frac{\sqrt 2 \pi f^{\rm VM}}{\sqrt{N_{\rm C}}}f(z)e^{-\om^2 r^2/2}.\nn
\lbqa{VMtrans2wf}
\begin{table}[!ht]
\center
\begin{tabular}{|c|c|c|c|c|}
\hline
&$\rh$&$\om$&$\ph$&$J/\Ps$\\
\hline\hline
$M_{\rm VM}[\GeV]$&0.770&0.782&1.019&3.097\\
\hline
$f^{\rm VM}[\GeV]$&0.153&0.0458&0.0791&0.270\\
\hline
$c_f^{\rm VM}$&$c_{u,d}= \pm1/\sqrt 2$&$c_{u,d}= +1/\sqrt 2 $&$c_s=1$&$c_c=1$\\
\hline
$\cN_{\rm long}(Q^2=0)$&15.10&15.47&15.70&19.03\\
\hline
$\om_{\rm long}^{-1}(Q^2=0)[\fm]$&0.597&0.658&0.536&0.290\\
\hline
$\cN_{\rm trans}(Q^2=0)$&6.75&7.61&7.59&9.05\\
\hline
$\om_{\rm trans}^{-1}(Q^2=0)[\fm]$&0.928&0.957&0.761&0.345\\
\hline
$\cN_{\rm long}(Q^2>1.6\GeV^2)$&15.10&15.47&15.70&19.03\\
\hline
$\om_{\rm long}^{-1}(Q^2>1.6\GeV^2)[\fm]$&0.597&0.658&0.536&0.290\\
\hline
$\cN_{\rm trans}(Q^2>1.6\GeV^2)$&11.50&13.21&11.37&9.05\\
\hline
$\om_{\rm trans}^{-1}(Q^2>1.6\GeV^2)[\fm]$&0.909&0.934&0.730&0.345\\
\hline
\end{tabular}
\caption{The parameters for the vectormeson wavefunctions}
\label{VMpara}
\end{table}

\end{document}